\documentclass[aps,twocolumn,prd,showpacs,nofootinbib]{revtex4}

\usepackage{bm}
\usepackage{latexsym}
\usepackage{amsfonts,amssymb,amsmath}
\usepackage{graphicx,epsfig}
\usepackage{psfrag}
\usepackage{subfigure}
\usepackage{ulem}
\usepackage{hyperref}
\usepackage{color}
\usepackage{appendix}
\usepackage{amsmath}
\usepackage{tikz}  % PERSONAL WARNING: requires to be compiled with TeXShop on the mac version, IAP distribution does not work 
\usepackage{float}

\hypersetup{
    bookmarks=true,         % show bookmarks bar?
    unicode=false,          % non-Latin characters in Acrobat’s bookmarks
    pdftoolbar=true,        % show Acrobat’s toolbar?
    pdfmenubar=true,        % show Acrobat’s menu?
    pdffitwindow=false,     % window fit to page when opened
    pdfstartview={FitH},    % fits the width of the page to the window
    pdftitle={My title},    % title
    pdfauthor={Author},     % author
    pdfsubject={Subject},   % subject of the document
    pdfcreator={Creator},   % creator of the document
    pdfproducer={Producer}, % producer of the document
    pdfkeywords={keyword1} {key2} {key3}, % list of keywords
    pdfnewwindow=true,      % links in new window
    colorlinks=true,       % false: boxed links; true: colored links
    linkcolor=red,          % color of internal links
    citecolor=cyan,        % color of links to bibliography
    filecolor=magenta,      % color of file links
    urlcolor=green,           % color of external links
    linktocpage=true
}

\newcommand{\dd}{\mathrm{d}}

\def\spose#1{\hbox to 0pt{#1\hss}}

\def\lta{\mathrel{\spose{\lower 3pt\hbox{$\mathchar"218$}}
     \raise 2.0pt\hbox{$\mathchar"13C$}}}
\def\gta{\mathrel{\spose{\lower 3pt\hbox{$\mathchar"218$}}
     \raise 2.0pt\hbox{$\mathchar"13E$}}}

\newcommand{\ie}{\textsl{i.e.~}}

\newcommand{\eg}{\textsl{e.g.~}}

\newcommand{\apriori}{\textsl{a priori~}}
\newcommand{\aposteriori}{\textsl{a posteriori~}}
\newcommand{\etc}{\textsl{etc.~}}

\newcommand{\Mp}{M_{_\mathrm{Pl}}}

\def\beq{\begin{equation}}
\def\eeq{\end{equation}}
\def\bea{\begin{eqnarray}}
\def\eea{\end{eqnarray}}
\def\eqref{\ref}

\newcommand{\sss}[1]{{\scriptscriptstyle{#1}}}

\newcommand{\uin}{\mathrm{in}}
\newcommand{\uend}{\mathrm{end}}
\newcommand{\umin}{\mathrm{min}}

\newcommand{\uc}{\mathrm{c}}

\newcommand{\uS}{\mathrm{S}}

\newcommand{\usssS}{\sss{\uS}}

\newcommand{\nS}{n_\usssS}

\newcommand{\xend}{x_\uend}
\newcommand{\xstar}{x_*}

\newcommand{\ee}{e}
\newcommand{\lo}{\sss{\mathrm{LO}}}
\newcommand{\nlo}{\sss{\mathrm{NLO}}}

\newcommand{\sr}{\sss{\mathrm{SR}}}\newcommand{\usr}{\sss{\mathrm{USR}}}
\newcommand{\hf}{\sss{\mathrm{HF}}}

\newcommand{\widthdouble}{8.9cm}
\newcommand{\widthsingle}{18.cm}

\begin{document}

%\title{Horizon-Flow versus Slow-Roll Analysis of Inflation}
%\title{Horizon-Flow Approach to Inflation}
\title{Horizon-Flow off-track for Inflation}

\author{Vincent Vennin} \email{vennin@iap.fr}
\affiliation{Institut d'Astrophysique de Paris, UMR 7095-CNRS,
Universit\'e Pierre et Marie Curie, 98bis boulevard Arago, 75014
Paris, France}

\date{\today}

\begin{abstract}
Inflation can be parametrized by means of truncated flow equations.  In this ``horizon-flow'' setup, generic results have been obtained, such as typical values for $r/\left(1-\nS\right)$. They are sometimes referred to as intrinsic features of inflation itself. In this paper we first show that the phenomenological class of inflationary potentials sampled by horizon flow is directly responsible for such predictions. They are therefore anything but generic. Furthermore, the horizon-flow setup is shown to rely on trajectories in phase space that differ from the slow roll. For a given potential, we demonstrate that this renders horizon flow blind to entire relevant inflationary regimes, for which the horizon-flow trajectory is shown to be unstable. This makes horizon flow a biased parametrization of inflation.
\end{abstract}

\maketitle

\section{Introduction}
\label{sec:intro}

Inflation is currently the leading paradigm for explaining the physical conditions that prevailed in the very early Universe~\cite{Starobinsky:1980te,Guth:1980zm,Linde:1981mu,Albrecht:1982wi,Linde:1983gd}. It describes a phase of accelerated expansion that solves the puzzles of the standard hot big bang model, and it provides a causal mechanism for generating inhomogeneities on cosmological scales~\cite{Mukhanov:1981xt,Mukhanov:1982nu,Starobinsky:1982ee,Guth:1982ec,Hawking:1982cz,Bardeen:1983qw}. These inhomogeneities result from the amplification of the unavoidable vacuum quantum fluctuations of the gravitational and matter fields during the accelerated expansion. In particular, inflation predicts that their spectrum should be almost scale invariant, with small deviations from scale invariance being related to the precise microphysics of inflation. This prediction is consistent with the current high precision astrophysical observations \cite{Hou:2012xq, Sievers:2013ica, Hinshaw:2012aka, Ade:2013rta}. In particular, the recent Planck measurement \cite{Ade:2013rta} of the cosmic microwave background temperature map gives together with WMAP polarization data a slightly red tilted scalar spectral index $\nS\simeq 0.96$, ruling out exact scale invariance $\nS=1$ at over $5\sigma$ and enabling us to constrain the inflationary models still allowed by the observations \cite{Martin:2013tda,Martin:2013nzq}.

Together with the absence of primordial non-Gaussianities and of isocurvature modes \cite{Ade:2013rta}, these results indicate that, at this stage, the full set of observations can be accounted for in the minimal setup, where inflation is driven by a single scalar field $\phi$, the inflaton field, minimally coupled to gravity, and evolving in some potential $V\left(\phi\right)$. The action for such a system is given by (hereafter $\Mp$ denotes the reduced Planck mass)
\begin{equation}
\label{eq:action}
S=\int{\left[\frac{\Mp^2}{2}R-\frac{1}{2}\partial_\mu\phi\partial^\mu\phi-V\left(\phi\right)\right]\sqrt{-g}\,\dd^4 x}\, ,
\end{equation}
where the background metric is chosen to be of the flat Friedmann-Lema\^itre-Robertson-Walker type, \ie the one of a homogeneous and isotropic expanding universe (about which fluctuations are evolved), given by $\dd s^2=-\dd t^2+a^2\left(t\right)\dd x^2$, where the scale factor $a\left(t\right)$ is a free function of time. However, the physical nature of the inflaton and its relation with the standard model of particle physics and its extensions remain elusive, since the inflationary mechanism is supposed to take place at very high energies in a regime where particle physics is not known and has not been tested in accelerators. Therefore the only requirement on $V$ is that it should be sufficiently flat to support inflation, but otherwise the multitude of inflaton candidates (with associated potentials) makes the theory as a whole hardly tractable, unless one restricts to a specific model. 

If one does so, within a given inflationary model $V(\phi)$, there exists a frame of approximation, the slow-roll approximation, which provides a set of manageable equations to calculate an attractor solution for the dynamics arising from the action~(\ref{eq:action}), and to consistently derive the statistical properties of cosmological perturbations produced during inflation. This is why, in order to constrain the inflationary scenario at a level matching the accuracy of the current data, a first approach is to scan the full set of models that have been proposed so far, and to test them one by one \cite{Martin:2013tda,Martin:2013nzq} making use of the slow-roll setup. 

Another strategy consists in developing model independent approaches and in studying generic parametrizations of inflation. Among these parametrizations is the ``horizon-flow'' setup~\cite{Hoffman:2000ue,Kinney:2002qn,Liddle:2003py,Ramirez:2005cy,Chongchitnan:2005pf} which relies on truncated flow equations describing the inflationary dynamics. The starting point is to define a set of flow parameters, based on time derivatives of the Hubble scale $H\equiv\dot{a}/a$ during inflation (a dot denoting a derivation with respect to cosmic time $t$), and to derive a set of equations for their variation in time. A finite subset of these equations is then solved numerically. Since all the observable quantities related to inflation directly depend on $H$ and the way it (slowly) varies with time, this is indeed a generic way to describe a full set of possible inflationary predictions. The goal of this paper is to investigate whether this horizon-flow approach can be used to robustly parametrize inflation.

Both approaches thus use a different function as an input: $V(\phi)$ for the slow-roll setup and $H(\phi)$ for the horizon-flow one. In section~\ref{sec:strategies} we review how these two strategies address the calculation of inflationary predictions. In particular, we point out that $H(\phi)$ and  $V(\phi)$ are explicitly related, and that the horizon-flow parametrization therefore only samples a particular set of inflationary potentials. In section~\ref{sec:SpecificPot}, we discuss the impact of restraining to such a class of phenomenological potentials, and we show that the typical values for $r/(1-\nS)$ that have been noticed in the literature are in fact in direct correspondence with the inflationary regimes supported by such potentials. However, examples are worked out where those relations break, showing that they are not generic. In section~\ref{sec:SpecificTraj}, we emphasize that the slow-roll and horizon-flow computational strategies differ by the phase space trajectory they respectively rest on, \ie the path the system follows in the $(\phi,\dot{\phi})$ plane as inflation proceeds in both setups. We carry out the slow-roll analysis of the potentials associated to the horizon-flow parametrization, computing the inflationary predictions for both trajectories and characterizing the discrepancies. We show that for a given potential, horizon flow does not sample all possible inflationary regimes, which introduces a bias in the way it parametrizes inflation. In some cases, its trajectory is even shown to be unstable. Finally in section~\ref{sec:conclusion}, we summarize our main results and conclude the discussion.
\section{Computing Inflationary Predictions}
\label{sec:strategies}
In this section we first recall how statistical properties of primordial cosmological fluctuations can be worked out in the framework of canonical single-field cosmological inflation~(\ref{eq:action}). We then review how the slow-roll and horizon-flow setups address the associated calculations.	
\subsection{The Single-Field Setup}
\label{sec:GenSetup}
In order to model the cosmological fluctuations, one needs to go beyond homogeneity and isotropy. When small fluctuations are added~\cite{Bardeen:1980kt,PPJPU} on top of the Friedman-Lema\^itre-Robertson-Walker metric introduced above and of the inflaton field, the scalar sector can be fully parametrized in terms of the Mukhanov-Sasaki variable $v$~\cite{Mukhanov:1981xt,Kodama:1985bj}. Expanding and varying the action~(\ref{eq:action}) at leading order in the perturbations, one can show that this gauge invariant quantity follows an equation of motion~\cite{Mukhanov:1990me} of the form
\beq
\label{eq:MSeom}
v_{\bm{k}}^{\prime\prime}+\left[k^2-\frac{\left(a\sqrt{\epsilon_1}\right)^{\prime\prime}}{a\sqrt{\epsilon_1}}\right]v_{\bm{k}}=0\, ,
\eeq
where $v_{\bm{k}}$ is the Fourier mode of $v$, and where $\epsilon_1\equiv 1-(a^\prime/a)^\prime/(a^\prime/a)^2$. Here, a prime denotes a derivative with respect to conformal time $\eta$, defined by $a\dd\eta=\dd t$. A similar equation can be obtained for tensor perturbations so that primordial gravity waves can be studied in the same way. Once Eq.~(\ref{eq:MSeom}) is solved, one can evaluate $v_{\bm{k}}$ at the end of inflation and calculate the power spectrum of curvature perturbations $\zeta=v/(a\sqrt{2\epsilon_1}\Mp)$ at that time, namely
\beq
\mathcal{P}_\zeta\left(\bm{k}\right)\equiv\frac{k^3}{2\pi^2}\left\vert\zeta_{\bm{k}}\right\vert^2=\frac{k^3}{4\pi^2a^2\epsilon_1}\left\vert v_{\bm{k}}\right\vert^2\, .
\eeq
To carry out such a program, two pieces of information are still missing. Firstly, one needs to set initial conditions for $v$ and $v^\prime$ at some reference time. A sensible choice of initial conditions is the Bunch-Davies vacuum where $v_{\bm{k}}\rightarrow \ee^{ik\eta}/\sqrt{2k}$ when $k/aH\rightarrow\infty$, which corresponds to setting each mode of the scalar perturbations in its Minkowski quantum ground state in the far sub-Hubble past. 

Secondly, one needs to specify the background function $a\sqrt{\epsilon_1}$. This is why as mentioned in the introduction,  at this point everything depends only on $a$ (or equivalently $H$) and the way it varies with time. This can be obtained as follows. From varying the action~(\ref{eq:action}) two dynamical equations arise for the background, namely the Friedmann equation, which equals the squared Hubble parameter to the energy density of the inflaton field, and the Klein-Gordon equation, which is the equation of motion of the field $\phi$. They are given by
\begin{eqnarray}
\label{eq:Friedman}
H^2=\frac{V+\dot{\phi}^2/2}{3\Mp^2}\, ,\\
\ddot{\phi}+3H\dot{\phi}+\frac{\dd V}{\dd \phi}=0\, .
\label{eq:KG}
\end{eqnarray}
Provided some initial conditions $\phi_\uin$ and $\dot{\phi}_\uin$, and assuming that the potential $V(\phi)$ is known, this system can be solved and the corresponding time evolution of $H$ (hence of $a$) can be inferred.  Then, Eq.~(\ref{eq:MSeom}) can be solved, and the statistical moments of $v$ can be calculated.

However, such a program is difficult to carry out in practice mainly because of three reasons. First, the system~(\ref{eq:Friedman})--(\ref{eq:KG}) and Eq.~(\ref{eq:MSeom}) can generally not be solved analytically. Second, a potential $V$ must be specified even if at these energies there is no unique candidate. Third, in general there is no obvious choice of initial conditions $\phi_\uin$ and $\dot{\phi}_\uin$. Both the slow-roll and horizon-flow approaches may simplify some of these issues. We now describe the strategies they rely on. 
\subsection{The Slow-Roll Approach}
\label{sec:SRsetup}
The slow-roll strategy relies on the assumption that the Hubble parameter and its time derivatives slowly vary with time during inflation, \ie that the deviation from de Sitter space-time is small.\footnote{Such an assumption is justified \aposteriori \eg by the fact that only small deviations from scale invariance are measured, with tight constraints on the level of gravity waves.} This can be characterized in terms of a hierarchy of ``slow-roll parameters.''

Although there are several possible sets of slow-roll parameters, in this paper, we choose to work with the Hubble-flow parameters $\{\epsilon_n\}$ defined by the flow equations \cite{Schwarz:2001vv,Schwarz:2004tz} 
\beq
\label{eq:epsFlow}
\epsilon_{n+1}=\frac{\dd\ln\vert\epsilon_n\vert}{\dd N}\, ,
\eeq
where the hierarchy is started at $\epsilon_0\equiv H_\uin/H$, and where $N\equiv \ln a$ is the number of $\ee$-folds. With this definition, all the $\epsilon_n$ are typically of the same order of magnitude. One has \textit{slow-roll} inflation as long as $\vert\epsilon_n\vert\ll 1$, for all $n>0$, while since $\epsilon_1=-\dot{H}/H^2=1-\ddot{a}/(aH^2)$, inflation ($\ddot{a}>0$) takes place provided $\epsilon_1<1$. Note that the definition of $\epsilon_1$ is of course consistent with the one introduced below Eq.~(\ref{eq:MSeom}). 

Now, when inflation is driven by a single scalar field, let us see how the system~(\ref{eq:Friedman})--(\ref{eq:KG}) gets simplified. Inserting the Klein-Gordon equation in the time derivative of the Friedman equation, one obtains $\dot{H}=-\dot{\phi}^2/(2\Mp^2)$, hence 
\beq
\label{eq:eps1exact}
\epsilon_1=-\frac{\dot{H}}{H^2}=3\frac{\dot{\phi}^2/2}{V(\phi)+\dot{\phi}^2/2}\, .
\eeq
The condition $\epsilon_1\ll 1$ thus implies that the kinetic energy of the inflaton is much smaller than its potential energy, namely $\dot{\phi}^2/2\ll V(\phi)$. Under this condition, the Friedmann equation simplifies and gives, at leading order in slow roll, $H^2\simeq V/(3\Mp^2)$.	

One can keep on and play the same game with $\epsilon_2$. Inserting the Klein-Gordon equation~(\ref{eq:KG}) in the time derivative of the relation $\dot{H}=-\dot{\phi}^2/(2\Mp^2)$ previously obtained, one gets $\ddot{H}=3H\dot{\phi}^2/\Mp^2+\dot{\phi}V^\prime/\Mp^2$, and
\beq
\epsilon_2=\frac{\ddot{H}}{H\dot{H}}-2\frac{\dot{H}}{H^2}=
6\left(\frac{\epsilon_1}{3}-\frac{V^\prime}{3H\dot{\phi}}-1\right)\, .
\eeq
Hereafter and unlike before, a prime denotes a derivative with respect to the field $\phi$. The condition $\epsilon_2\ll 1$ thus implies that, at leading order in slow roll, $\dot{\phi}\simeq-V^\prime/(3H)$, which means that the acceleration term can be neglected in the Klein-Gordon equation. This is particularly interesting since it lowers by $1$ the order of the differential equation satisfied by $\phi$. As a consequence, it removes dependence on the initial conditions by singling out a specific trajectory, and analytical solutions are available in most cases. In this manner it solves the first and the third difficulties mentioned at the end of section~\ref{sec:GenSetup}.

More explicitly, since $\dd N=H \dd t$, at leading order in slow roll the Klein-Gordon equation reads $\dd N=-3H^2\dd\phi/V^\prime$. Plugging in the slow-roll leading order of the Friedman equation $H^2_{\sr,\lo}=V/(3\Mp^2)$, one obtains
\beq
\label{eq:sr:trajlo}
\Delta N^{\sr,\lo}=-\frac{1}{\Mp^2}\int_{\phi_\uin}^{\phi_\uend}\frac{V}{V^\prime}\dd\phi\, ,
\eeq
where $\Delta N\equiv N_\uend-N_\uin$, $\phi_\uin$ is the value of $\phi$ at some initial time $N_\uin$, and $\phi_\uend$ is the value of $\phi$ at some final time $N_\uend$. This represents the leading order ($\mathrm{LO}$) of the slow-roll (SR) trajectory. Inverting this relation yields the value of $\phi$ at any time $N$.

Furthermore, it turns out that the slow-roll trajectory is a powerful attractor~\cite{Remmen:2013eja} of the inflationary dynamics, that is to say, starting from a large basin of possible initial conditions $\phi_\uin$ and $\dot{\phi}_\uin$, the system quickly converges towards the slow-roll trajectory. We will come back to this point in section~\ref{sec:SpecificTraj}, but we can already notice that it is a strong physical motivation to work within the slow-roll framework.

It is also interesting to remark that under the slow-roll approximation, the slow-roll hierarchy can be easily expressed in terms of $V$ and its derivatives. Indeed, starting from $H^2\simeq V/(3\Mp^2)$ the derivative relation $\dd/\dd t=\dot{\phi}\,\dd/\dd\phi\simeq-V^\prime/(3H)\dd/\dd\phi$ gives rise to
\beq
\left.\frac{\dd}{\dd N}\right\vert^{\sr,\lo}=-\Mp^2\frac{V^\prime}{V}\frac{\dd}{\dd\phi}\, .
\eeq
Repeatedly applying this identity, one obtains, at leading order in slow roll,
\bea
\label{eq:srhierarchySRlostart}
\label{eq:srhierarchySReps0}
\epsilon_0^\lo&=&H_\uin\sqrt{\frac{3\Mp^2}{V}}\, ,\\
\label{eq:srhierarchySReps1}
\epsilon_1^\lo&=&\frac{\Mp^2}{2}\left(\frac{{V^\prime}}{V}\right)^2\, ,\\
\label{eq:srhierarchySReps2}
\epsilon_2^\lo&=&2\Mp^2\left[\left(\frac{V^\prime}{V}\right)^2-\frac{V^{\prime\prime}}{V}\right]\, ,\\
\label{eq:srhierarchySRloend}
\label{eq:srhierarchySReps3}
\epsilon_3^\lo&=&\frac{2\Mp^4}{\epsilon_2^\lo}\left[ \frac{V^{\prime\prime\prime}V^\prime}{V^2}-3\frac{V^{\prime\prime}{V^\prime}^2}{V^3}+2\left(\frac{V^\prime}{V}\right)^4\right]\, ,\\
& &\nonumber
\eea
and the following slow-roll parameters can be computed in the same way.

Since the slow-roll parameters entirely characterize the time evolution of $H$ (and of $a$), it is now obvious that the solutions to Eq.~(\ref{eq:MSeom}) can be expressed in terms of them, hence the statistical moments of cosmological fluctuations at the end of inflation too. For example, at leading order in slow roll, the scalar power spectrum is given by
\beq
k^3\mathcal{P}_\zeta=\frac{H_*^2}{8\pi^2\Mp^2\epsilon_{1*}}\left[1-(2\epsilon_{1*}+\epsilon_{2*})\ln\frac{k}{k_\mathrm{P}}+\cdots\right]\, ,
\eeq
where a star means that quantities must be evaluated at the Hubble exit time of some pivot scale $k_\mathrm{P}$ of astrophysical interest today. One can see that the scalar power spectrum is scale invariant, with logarithmic corrections whose amplitude is slow roll suppressed. They can be described in terms of the spectral index 
\beq
\label{eq:ns:srlo}
\nS\equiv 1+\frac{\dd\ln \mathcal{P}_\zeta}{\dd\ln k}\simeq 1-2\epsilon_{1*}-\epsilon_{2*}\, ,
\eeq
the last expression being given at leading order in slow roll. As already mentioned, the same program can be carried out for tensor modes and the power spectrum of gravity waves $\mathcal{P}_h$ can be obtained in the same manner. The ratio $r$ of its amplitude to the scalar power spectrum amplitude is often used to characterize the primordial level of gravity waves. At leading order in slow roll, one has 
\beq
\label{eq:r:srlo}
r\equiv\frac{\mathcal{P}_h}{\mathcal{P}_\zeta}\simeq 16\epsilon_{1*}\, .
\eeq

The slow-roll program is therefore straightforward. The Hubble crossing time of the pivot scale depends on the subsequent thermal history of the Universe, and is typically located $\Delta N_*\simeq 50$ $\ee$-folds before the end of inflation. Given a potential $V(\phi)$, one thus integrates the slow-roll trajectory~(\ref{eq:sr:trajlo}) $\Delta N_*$ $\ee$-folds prior to the end of inflation (defined as $\epsilon_1=1$) and evaluates the potential and its derivatives there. Making use of Eqs.~(\ref{eq:srhierarchySRlostart})--(\ref{eq:srhierarchySRloend}), the slow-roll parameters ${\epsilon_{n*}}$ are obtained, and physical quantities such as $\nS$ and $r$ can be computed by means of the formulas~(\ref{eq:ns:srlo}) and (\ref{eq:r:srlo}).

In the following it will turn useful to make use of next-to-leading order ($\mathrm{NLO}$) expressions in slow roll, \ie one order further than above. This is why we end this section by deriving such formulas.  The starting point is to combine Eqs.~(\ref{eq:Friedman}) and (\ref{eq:eps1exact}) into
\beq
\label{eq:HVeps1}
H^2=\frac{V}{3\Mp^2}\left(1-\frac{\epsilon_1}{3}\right)^{-1}\, .
\eeq
Together with the Friedman equation~(\ref{eq:Friedman}), this gives rise to $\dot{\phi}^2=2V\epsilon_1/(3-\epsilon_1)$. These two formulas enable us to recast $\dd N=H\dd\phi/\dot{\phi}$ as
\beq
\label{eq:dNdphieps1}
\dd N=\pm\frac{1}{\Mp}\frac{\dd\phi}{\sqrt{2\epsilon_1}}\, .
\eeq
From here the slow-roll parameters at next-to-leading order can be obtained as follows. Rewriting Eq.~(\ref{eq:HVeps1}) as $\epsilon_0=\epsilon_0^\lo\sqrt{1-\epsilon_1/3}$, and iteratively applying 
\beq
\left.\frac{\dd}{\dd N}\right\vert^{\sr,\nlo}=\sqrt{\frac{\epsilon_1^\nlo}{\epsilon_1^\lo}}\left.\frac{\dd}{\dd N}\right\vert^{\sr,\lo}
\eeq
which comes from Eq.~(\ref{eq:dNdphieps1}), one obtains an expression for the slow-roll parameters at next-to-leading order in terms of the slow-roll parameters at leading order, which read
\bea
\label{eq:srhierarchySRnlostart}
\label{eq:srhierarchySRnloeps0}
\epsilon_0^\nlo&=&\epsilon_0^\lo\left(1-\frac{\epsilon_1^\lo}{6}\right)\, ,\\
\label{eq:srhierarchySRnloeps1}
\epsilon_1^\nlo&=&\epsilon_1^\lo\left(1-\frac{\epsilon_2^\lo}{3}\right)\, ,\\
\label{eq:srhierarchySRnloeps2}
\epsilon_2^\nlo&=&\epsilon_2^\lo\left(1-\frac{\epsilon_2^\lo}{6}-\frac{\epsilon_3^\lo}{3}\right)\, ,\\
\label{eq:srhierarchySRnloend}
\label{eq:srhierarchySRnloeps3}
\epsilon_3^\nlo&=&\epsilon_3^\lo\left(1-\frac{\epsilon_2^\lo}{3}-\frac{\epsilon_4^\lo}{3}\right)\, ,\\
& &\nonumber
\eea
where the following slow-roll parameters can be computed in the same manner, and where the slow-roll parameters at leading order in the right-hand sides are given by Eqs.~(\ref{eq:srhierarchySRlostart})--(\ref{eq:srhierarchySRloend}). If one wanted to keep on and go up to next-to-next-to-leading order, one would proceed in exactly the same way, but here it is enough to stop at next-to-leading order.

Let us move on to the slow-roll trajectory. At next-to-leading order, it proceeds from combining Eq.~(\ref{eq:dNdphieps1}) and Eq.~(\ref{eq:srhierarchySRnloeps1}), which gives rise to
\beq
\label{eq:trajSRnlo}
\Delta N^{\sr,\nlo}=-\frac{1}{\Mp^2}\int_{\phi_\uin}^{\phi_\uend}\frac{V}{V^\prime}\dd\phi+\frac{1}{3}\ln\left(\frac{V^\prime_\uend/V_\uend}{V^\prime_\uin/V_\uin}\right)\, .
\eeq
At last, at next-to-leading order in slow roll, the spectral index and the tensor to scalar ratio are given by~\cite{Gong:2001he,Martin:2013uma}
\bea
\nS &=&1-2\epsilon_{1*}-\epsilon_{2*}-2\epsilon_{1*}^2
\nonumber\\ & & 
\label{eq:ns:srnlo}
-(2C+3)\epsilon_{1*}\epsilon_{2*}-C\epsilon_{2*}\epsilon_{3*}\, ,\\
r&=&16\epsilon_{1*}\left(1+C\epsilon_{2*}\right)\, ,
\label{eq:r:srnlo}
\eea
where $C\equiv\gamma_\mathrm{E}+\ln 2-2\simeq -0.7296$, $\gamma_\mathrm{E}$ being the Euler constant. 
\subsection{The Horizon-Flow Approach}
\label{sec:hfintro}
Contrary to the slow-roll approach which consists in solving the system~(\ref{eq:Friedman})--(\ref{eq:KG}) with some approximation, the horizon-flow strategy~\cite{Hoffman:2000ue,Kinney:2002qn,Liddle:2003py,Chongchitnan:2005pf,Ramirez:2005cy} uses flow equations of the kind~(\ref{eq:epsFlow}) as the fundamental input to derive physical observables such as $\nS$ or $r$. In this section we review how this can be achieved.

A first remark is that for single-field inflation, the flow parameters can be cast as functions of $\phi$ instead of $t$ in full generality. Indeed, since the inflaton field $\phi$ varies during inflation under the effect of its potential and initial speed, $\phi$ can be used as a time label itself, which is unambiguous provided $\phi$ is monotonic in time. Concretely, identifying the expression $\dot{H}=-\dot{\phi}^2/(2\Mp^2)$ found above Eq.~(\ref{eq:eps1exact}) with the simple relation $\dot{H}=H^\prime\dot{\phi}$, one obtains
\begin{equation}
\label{eq:phidotHprime}
\dot{\phi}=-2\Mp^2H^\prime\, .
\end{equation}
This enables us to relate the derivative with respect to the number of $\ee$-folds to the derivative with respect to the inflaton field, $\dd/\dd N=\dot{\phi}H^{-1}\dd/\dd\phi$, by
\beq
\label{eq:dNdphi}
\frac{\dd}{\dd N}=-2\Mp^2\frac{H^\prime}{H}\frac{\dd}{\dd\phi}\, .
\eeq
The slow-roll hierarchy $\{\epsilon_n\}$ can thus be expressed only in terms of $H(\phi)$ and its derivatives. Starting from $\epsilon_0=H_\uin/H$ and repeatedly applying Eq.~(\ref{eq:dNdphi}), the flow equations~(\ref{eq:epsFlow}) give rise to
\bea
\label{eq:srhierarchystart}
\label{eq:hf:eps0}
\epsilon_0&\equiv&\frac{H_\uin}{H}\, ,\\
\label{eq:hf:eps1}
\epsilon_1&\equiv&\frac{\dd\ln\left\vert\epsilon_0\right\vert}{\dd N}=2\Mp^2\left(\frac{{H^\prime}}{H}\right)^2\, ,\\
\label{eq:hf:eps2}
\epsilon_2&\equiv&\frac{\dd\ln\left\vert\epsilon_1\right\vert}{\dd N}=4\Mp^2\left[\left(\frac{H^\prime}{H}\right)^2-\frac{H^{\prime\prime}}{H}\right]\, ,\\
\label{eq:srhierarchyend}
\label{eq:hf:eps3}
\epsilon_3&\equiv&\frac{\dd\ln\left\vert\epsilon_2\right\vert}{\dd N}=2\Mp^2\left[2\left(\frac{H^\prime}{H}\right)^2+\frac{H^{\prime\prime\prime}}{H^\prime}-3\frac{H^ {\prime\prime}}{H}\right]\times\nonumber\\& &\left(1-\frac{HH^{\prime\prime}}{{H^\prime}^2}\right)^{-1}\, ,\\
& &\nonumber
\eea
and the following parameters can be iteratively computed in the same manner. Note that contrary to Eqs.~(\ref{eq:srhierarchySRlostart})--(\ref{eq:srhierarchySRloend}), all the above expressions are exact and do not rely on any kind of approximation.
It is also clear that the slow-roll parameters depend only on $H(\phi)$ and its derivatives, and that this function therefore contains all the relevant information to derive the physical predictions of inflation. 

In the horizon-flow literature~\cite{Hoffman:2000ue,Kinney:2002qn,Liddle:2003py,Chongchitnan:2005pf,Ramirez:2005cy} a different set of flow parameters is often used, which leads the way to a computational program that we now explain. In this set of parameters, $\epsilon_1$ and $\epsilon_2$ are supplemented with~\cite{Kinney:2002qn}
\beq
\label{eq:lambdaDef}
{}^l\lambda_H=\left(2\Mp^2\right)^l\frac{\left(H^\prime\right)^{l-1}}{H^l}\frac{\dd^{l+1}H}{\dd\phi^{l+1}}\, ,\quad \mathrm{for}\ l>1\, . 
\eeq
From here\footnote{These parameters ${}^l\lambda_H$ are related to the parameters ${}^l\beta_H$ defined in Ref.~\cite{Liddle:1994dx} by ${}^l\lambda_H=\left({}^l\beta_H\right)^l$.} a set of flow equations similar to Eq.~(\ref{eq:epsFlow}) can be derived: $\dd\epsilon_1/\dd N=\epsilon_1\epsilon_2$, $\dd\epsilon_2/\dd N= 2 \left({}^2 \lambda_H\right)- 2\epsilon_1^2 +3\epsilon_1 \epsilon_2$, and
\beq
\label{eq:lambdaFlow}
\frac{\dd{}^l\lambda_H}{\dd N}=\left(\frac{l-1}{2}\epsilon_2+\epsilon_1\right){}^l\lambda_H-{}^{l+1}\lambda_H\, .
\eeq
One should note that contrary to the hierarchy $\{\epsilon_n\}$, these flow parameters are of increasing order in slow roll. Obviously both hierarchies are explicitly related.

One way to solve the infinite system~(\ref{eq:lambdaFlow}) is to truncate it at some level, by setting all flow parameters beyond a sufficiently high order in the hierarchy to zero, \ie ${}^{l}\lambda_H=0$ for $l>M$, where $M$ is a suitably large integer (in the literature~\cite{Kinney:2002qn,Liddle:2003py,Chongchitnan:2005pf,Ramirez:2005cy}, $M=5$ has essentially been investigated). The flow equations then comprise a closed finite set. 
Once initial conditions on the flow parameters $\epsilon_1$, $\epsilon_2$, ${}^2\lambda_H$, $\cdots$, ${}^M\lambda_H$ are chosen, the horizon-flow computational program consists in integrating the flow equations~(\ref{eq:lambdaFlow}) forward in time until one of the three following scenarios occurs: 
\begin{description}
\item[(i)] The parameter $\epsilon_1$ reaches $1$ and inflation naturally ends. From here the flow equations are integrated $\Delta N_*$ $\ee$-folds backward in time and the observables are calculated there.
\item[(ii)] The system reaches a late-time fixed point, where observables are calculated.
\item[(iii)] None of this happens: inflation never ends (after a ``long'' integration time, typically $1000$ $\ee$-folds) and no fixed point is reached. In this case the model is just thrown away.
\end{description}
Note that the predictions are computed thanks to the slow-roll approximated formulas~(\ref{eq:ns:srlo}) and (\ref{eq:r:srlo}) or (\ref{eq:ns:srnlo}) and (\ref{eq:r:srnlo}), expressed in terms of the chosen set of flow parameters.

Then one proceeds with running the same algorithm again, with different values of initial flow parameters and $\Delta N_*$, so on and so forth, until a huge number of predictions are computed among which ``typical'' features are searched for. The parameters ($\Delta N_*$ and initial flow parameters) are usually drawn in predefined ranges of values, the priors. The width of the prior intervals for the initial flow parameters is usually reduced by some factor (typically $5$~\cite{Chongchitnan:2005pf} or $10$~\cite{Kinney:2002qn,Ramirez:2005cy}) for each higher order in the hierarchy.

A crucial remark, made in Ref.~\cite{Liddle:2003py}, is that truncating the hierarchy $\{{}^l\lambda_H\}$ at some order $M$ is actually equivalent to requiring that $\dd^{M+2}H/\dd\phi^{M+2}$ vanishes, which means that $H(\phi)$ must be a polynomial function of order $M+1$
\beq
\label{eq:Hpolynomial}
H(\phi)=H_0\left[1+\sum_{i=1}^{M+1}a_i\left(\frac{\phi}{\Mp}\right)^i\right],
\eeq
the $a_i$ coefficients being directly related to the initial flow parameters of the computational algorithm detailed above. Inflation is thus described in terms of a model depending on $M+1$ free parameters (the $a_i$, or equivalently, initial values for $\epsilon_1$, $\epsilon_2$, ${}^2\lambda_H$, $\cdots$, ${}^M\lambda_H$), on which a prior range of variation is set. The dependence on the choice of such priors, and on the parameter set one uses (initial flow parameters ${}^l\lambda_H$, $a_i$ parameters, or other possible choices), is investigated in Ref.~\cite{Ramirez:2005cy}. It is shown that while there remains some concentration of points around the above mentioned fixed points under the different parameter sets, there is significant variation in the predictions among them.

Let us insist that in this setup, inflation is parametrized by a free generic function $H(\phi)$, and that it is also the case in the more common approach where one solves Eqs.~(\ref{eq:Friedman}) and (\ref{eq:KG}) with some function $V\left(\phi\right)$. Interestingly enough, it turns out~\cite{Liddle:2003py} that the two functions are straightforwardly related. Indeed, plugging the relation~(\ref{eq:phidotHprime}) in the Friedman equation~(\ref{eq:Friedman}), one obtains
\begin{equation}
\label{eq:VversusH}
V=3\Mp^2H^2-2\Mp^4{H^\prime}^2\, .
\end{equation}
Therefore, horizon flow does not really solve the second difficulty mentioned at the end of section~\ref{sec:GenSetup} (\ie the necessity to specify a potential) since it implicitly assumes a specific potential, through the choice of $H$, and a specific initial value $\phi_\uin$ through the choice of the initial flow parameters. 

Moreover, the potential $V(\phi)$ derives in principal from the physical origin of the inflaton field, and the free parameters it contains are usually related to physical quantities such as charges, coupling constants, masses, \etc Therefore it may seem more sensible and physically appealing to parametrize inflation in terms of these quantities (and to choose corresponding simple priors on them), instead of using the integration constants of the flow equations, which \apriori do not carry any particular physical meaning. 

In passing, let us note that the horizon-flow computational program has also been used as a potential reconstruction technique~\cite{Easther:2002rw, Peiris:2003ff, Kinney:2003uw, Chen:2004nx, Powell:2007gu, Powell:2008bi, Contaldi:2013mua}.\footnote{In Ref.~\cite{Powell:2008bi}, note that the horizon-flow setup is extended to noncanonical single-field models with varying speed of sound $c_s$, the inverse of which is parametrized by a truncated Taylor expansion of the type~(\ref{eq:Hpolynomial}), with associated $c_s$-flow equations of the type~(\ref{eq:lambdaFlow}).}
A selection rule is added to the algorithm detailed previously that specifies an admitted region in observable parameter space (usually defining central values for $\nS$ and $r$ with associated error bars). When a trajectory is integrated, its predictions are computed and the trajectory is kept only if these predictions lie in the admitted region. For all the remaining trajectories at the end of the program, the potential is computed using Eq.~(\ref{eq:VversusH}) and all the potentials are superimposed on a single plot to see which typical shape comes out. Obviously, such an approach to potential reconstruction suffers from the same shortcomings discussed in this paper as horizon flow itself. 

In the two next sections, we briefly review the two main results of this paper: the origin of the so-called ``typical'' predictions of horizon-flow inflation, and the bias introduced by horizon-flow trajectories in the parametrization of single-field inflation.
\subsection{``Typical'' Predictions}
\label{sec:TypPred}
In the references mentioned above two typical denser regions turn out to be sampled: either $r_{16}/(1-\nS)=1/2$ or $r_{16}=0$ (where $r_{16}=r/16$ corresponds to the ``$r$'' parameter defined in Refs.~\cite{Kinney:2002qn,Ramirez:2005cy}). Actually, this can be understood with the following heuristic argument. The first order slow-roll relations $r=16\epsilon_1$ and $\nS-1=-2\epsilon_1-\epsilon_2$, combined with the flow equations~(\ref{eq:epsFlow}), allow one to express the number of $\ee$-folds derivatives of $\nS$ and $r$ in terms of $\nS$, $r$ and $\epsilon_3$. Working with the two variables $s\equiv -r/8+(1-\nS)$ and $r$ instead of $\nS$ and $r$, one obtains
\bea
\frac{\dd s}{\dd N}&=&\epsilon_3 s\, ,\\
\frac{\dd r}{\dd N}&=& r s\, .
\eea
If the $\lbrace\epsilon_n\rbrace$ hierarchy is truncated at $n=4$ (\ie $\epsilon_3$ is constant and $\epsilon_{n>3}$ vanish), this system contains two fixed points: either $\epsilon_3=0$ and $r=0$, which leads to $r_{16}/(1-\nS)=0$, or $s=0$, which by definition leads to $r_{16}/(1-\nS)=1/2$. This exactly corresponds to the denser regions mentioned above and  matches the early results of Refs.~\citep{Hoffman:2000ue} (be careful that another normalization is again used in this paper, where $T/S=10\, r_{16}=5r/8$).  If this were concluded to be generic predictions of inflation, this would have important consequences for inflation itself, since \eg the region $r_{16}/(1-\nS)=1/2$ is now strongly disfavored by the most recent observations~\cite{Ade:2013rta}.

In Ref.~\cite{Kinney:2002qn} these fixed points are shown to be generic fixed points of the hierarchy~(\ref{eq:lambdaDef}) at any order (\ie for any $M$) and their stability is studied in Ref.~\cite{Chongchitnan:2005pf}. However in Refs.~\cite{Kinney:2002qn,Ramirez:2005cy} it is also noticed that even if the numerical models generated by the above algorithm cluster not far from the region $r_{16}/(1-\nS)=1/2$, a better fit is given by
\beq
\frac{r_{16}}{1-\nS}\simeq\frac{1}{3}\, .
 \eeq
In this paper we puzzle out this discrepancy for the first time, analytically showing where this number $1/3$ comes from. Indeed, Eq.~(\ref{eq:VversusH}) shows that when using a parametrization of the form (\ref{eq:Hpolynomial}), only a particular set of inflationary potentials is actually investigated, namely polynomial potentials with some relations among the coefficients. In section~\ref{sec:SpecificPot} we discuss the impact of restraining to such a class of phenomenological models, and we show that the relations $r_{16}/(1-\nS)\sim 1/3$ and $r_{16}\sim 0$ actually correspond to the different inflationary regimes of such potentials. 
\subsection{Inflationary Trajectories}
\label{sec:Traj}
Even if Eq.~(\ref{eq:VversusH}) explicitly relates $H$ and $V$, the corresponding horizon-flow and slow-roll analyses are different because they rely on different inflationary trajectories. In this section we first explain why it is so, before we investigate the consequences of this difference.
\subsubsection{Why Horizon-Flow and Slow-Roll trajectories differ}
Since the system~(\ref{eq:Friedman})--(\ref{eq:KG}) is second order in time derivative, its solutions form a one-dimensional set of inflationary trajectories, \ie an infinite bundle of paths in phase space $(\phi,\dot{\phi})$ (examples are displayed and commented on in section~\ref{sec:HFversusSRtraj}). As we shall now see, both the slow-roll and horizon-flow setups rely on a single trajectory each, and do not scan this whole set of possible dynamics.

The trajectory on which the slow-roll approach rests has already been explicated in section~\ref{sec:SRsetup}, see Eq.~(\ref{eq:sr:trajlo}) for its leading order expression and Eq.~(\ref{eq:trajSRnlo}) for its next-to-leading order expression. Even if a complete form can only be attained asymptotically by a perturbative calculation, it is nonetheless a well-defined and unique object. One should therefore be aware of the subtlety that ``slow roll'' both refers to a perturbative computational framework and to a specific inflationary trajectory. The latter is calculable by the former, and is known to be a powerful attractor~\cite{Remmen:2013eja} of the inflationary dynamics. This is why it makes sense to study inflation along its line.

On the other hand, the horizon-flow formalism also implies a particular inflationary trajectory. It does not explicitly make use of it, which is why it has not really been noticed in the literature so far, but such a trajectory is implicitly contained in the computational approach of horizon flow. Indeed, since the $H$ function is defined through Eq.~(\ref{eq:Friedman}) on the full phase space $(\phi,\dot{\phi})$, reducing it to an $H(\phi)$ function only
\beq
H(\phi,\dot{\phi})\rightarrow H(\phi)
\eeq
necessarily implies some relation $\dot\phi(\phi)$, that is, by definition, a trajectory. It is actually given by Eq.~(\ref{eq:phidotHprime}). More precisely, the trajectory associated to some $H^\hf(\phi)$ function is basically given by Eq.~(\ref{eq:dNdphi}), \ie
\beq
\label{eq:hf:traj}
\Delta N^\hf=-\frac{1}{2\Mp^2}\int_{\phi_\uin}^{\phi_\uend}\frac{H^\hf}{\left(H^\hf\right)^\prime}\dd\phi\, ,
\eeq
which is exact and does not rely on any approximation, and where ``$\mathrm{HF}$'' stresses that we are working within the horizon-flow setup.

The problem can therefore be formulated as follows. Starting from an $H^\hf(\phi)$ function [typically Eq.~(\ref{eq:Hpolynomial})], horizon flow consists of studying inflation along the trajectory~(\ref{eq:hf:traj}). Now, thanks to Eq.~(\ref{eq:VversusH}), a potential $V$ can be associated to $H^\hf$, so that the slow-roll analysis can be worked out in this potential, and inflation can be studied along the slow-roll trajectory. The question is whether these two trajectories match or not. 

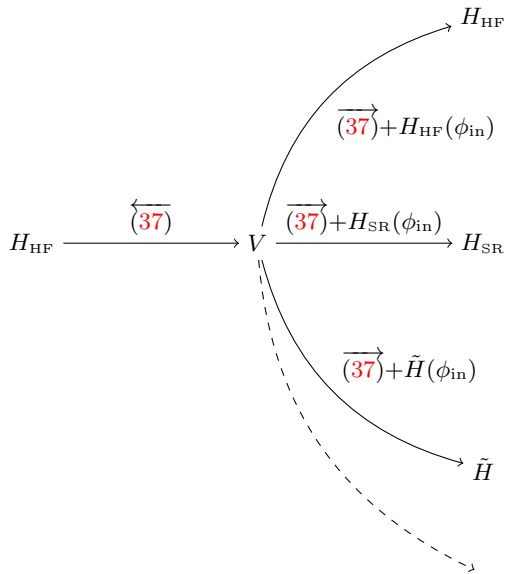
\begin{figure}[t]
\begin{center}
\begin{tikzpicture}[node distance=3cm, auto]
  \node (V) {$V$};
  \node (Hhf) [left of=V] {$H_\hf$};
  \node (Hsr) [right of=V] {$H_\sr$};
  \node (Hhfr) [above of=Hsr] {$H_\hf$};
  \node (H) [below of=Hsr] {$\tilde{H}$};
  \node (H2) [below of=H, node distance=1.4cm] {};
  \draw[->] (Hhf) to node {$\overleftarrow{\mathrm{(\ref{eq:VversusH})}}$} (V);
  \draw[->, bend left] (V) to node [swap] {$\overrightarrow{\mathrm{(\ref{eq:VversusH})}}$+$H_\hf(\phi_\uin)$} (Hhfr);
  \draw[->] (V) to node {$\overrightarrow{\mathrm{(\ref{eq:VversusH})}}$+$H_\sr(\phi_\uin)$} (Hsr);
  \draw[->, bend right] (V) to node {$\overrightarrow{\mathrm{(\ref{eq:VversusH})}}$+$\tilde{H}(\phi_\uin)$} (H);
  \draw[->, bend right, dashed] (V) to node {} (H2);
\end{tikzpicture}
\caption{Relations between $H$ and $V$ functions. Starting from a given $H_\hf$ function, the associated potential $V$ can be obtained using Eq.~(\ref{eq:VversusH}) from the right-hand side to the left-hand side [hence the direction of the arrow above ${\mathrm{(\ref{eq:VversusH})}}$]. Starting now from this potential $V$, several $H$ functions can be obtained using Eq.~(\ref{eq:VversusH}) from the left-hand side to the right-hand side (\ie solving a first order differential equation, that involves one integration constant), depending on the initial condition $H(\phi_\uin)$ one chooses. The one corresponding to slow roll, obtained by setting  $H(\phi_\uin)=H_\sr(\phi_\uin)$, has \apriori no reason to mach the initial  $H_\hf$ function corresponding to $H(\phi_\uin)=H_\hf(\phi_\uin)$.}
\label{fig:sketch}
\end{center}
\end{figure}
In general they do not for the following reason. Thanks to the Friedman equation~(\ref{eq:Friedman}), let us recall that the Hubble parameter $H$ is a function defined along any trajectory supported by a given potential $V$. All these $H$ functions are different, technically because  they correspond to solutions of Eq.~(\ref{eq:VversusH}) (viewed as a differential equation giving $H$ once $V$ is fixed) with different initial conditions $H(\phi_\uin)$. Among these functions is the one corresponding to $H^\hf$ if one chooses $H^\hf(\phi_\uin)$ as an initial condition, but one can also find $H^\sr$ which corresponds to the slow-roll trajectory if one chooses $H^\sr(\phi_\uin)$ as an initial condition, or any other $\tilde{H}$ function corresponding to any other trajectory and associated initial condition. The sketch displayed in Fig.~\ref{fig:sketch} summarizes the situation.
Since there is no reason why $H^\hf(\phi_\uin)=H^\sr(\phi_\uin)$ \apriori, the two functions $H^\hf$ and $H^\sr$ are different; hence the slow-roll trajectory
\beq
\Delta N^\sr=-\frac{1}{2\Mp^2}\int_{\phi_\uin}^{\phi_\uend}\frac{H^\sr}{\left(H^\sr\right)^\prime}\,\dd\phi
\eeq
differs from the horizon-flow one~(\ref{eq:hf:traj}) in general. 

\begin{figure*}[t]
\begin{center}
\includegraphics[width=\widthsingle]{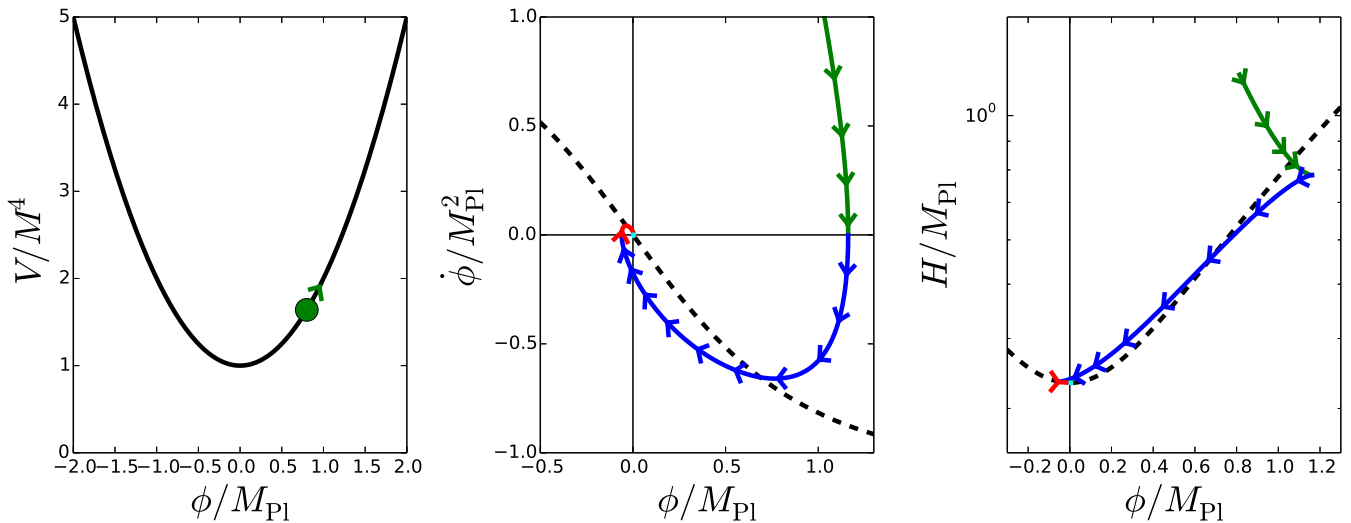}
\caption{Left panel: Potential $V/M^4=1+(\phi/\Mp)^2$ as a function of $\phi$. The green disk and arrow stand for the initial value and sign of velocity of the inflaton field for the trajectory displayed in the middle and right panels. Middle panel: Numerical integration of equation~(\ref{eq:KG}) from $\phi_\uin/\Mp=0.8$ and $\dot{\phi}_\uin/\Mp^2=2$ (colored line), displayed in the phase plane $(\phi,\dot{\phi})$. Each color corresponds to a different piece of it, with alternate signs of $\dot{\phi}$. The black dashed line stands for the slow-roll leading order trajectory $\dot{\phi}=-V^\prime/(3 H^{\sr,\lo})$. Right panel: Hubble parameter $H(\phi,\dot{\phi})$ evaluated along this trajectory (same color code) with Eq.~(\ref{eq:Friedman}). Note that the logarithmic scale is used for $H$ for display convenience. Again, the black dashed line stands for the slow-roll leading order solution $H^{\sr,\lo}(\phi)=V(\phi)/(3\Mp^2)$. In the middle and right panels, the arrows indicate in which direction inflation proceeds.}
\label{fig:TrajEx}
\end{center}
\end{figure*}

This being said, since the horizon-flow algorithm imposes that we start from small values of the flow parameters $\epsilon^\hf$, and since the slow-roll trajectory is in any case an attractor, the departure from slow roll is initially small and should remain so. This is why at first sight, one may claim that predictions should not be too affected since when the slow-roll conditions ($\epsilon\ll 1$) are verified for $H^\sr$ and $H^\hf$, the differentials in the predictions of both frames are slow roll suppressed quantities (and one needs to use next-to-leading order expressions in slow roll to consistently compare them,\footnote{A physical quantity $\mathcal{P}$ computed in the horizon-flow parametrization $\mathcal{P}_\hf$ only differs from the slow-roll one $\mathcal{P}_\sr$ by slow roll suppressed quantities; that is $\mathcal{P}_\hf=\mathcal{P}_\sr[1+\mathcal{O}(\epsilon)+\cdots]$, where $\epsilon$ stands for first order terms in slow roll. If $\mathcal{P}_\sr$ is computed in the slow-roll frame of approximation, $\mathcal{P}_\sr=\mathcal{P}_\sr^\lo[1+\tilde{\mathcal{O}}(\epsilon)+\cdots]$, one has at leading order $\mathcal{P}_\hf=\mathcal{P}_\sr^\lo[1+\tilde{\mathcal{O}}(\epsilon)+\mathcal{O}(\epsilon)+\cdots]$. To consistently derive the leading order differential $\mathcal{P}_\hf-\mathcal{P}_\sr=\mathcal{P}_\sr^\lo\mathcal{O}(\epsilon)+\cdots$, the term $\propto\tilde{\mathcal{O}}(\epsilon)$ must therefore be computed; \ie the slow-roll quantities must be worked out at next-to-leading order.} as in section~\ref{sec:trajPredictions}). 

For example, from deriving Eq.~(\ref{eq:VversusH}) with respect to $\phi$, one can rewrite Eq.~(\ref{eq:hf:traj}) as
\beq
\label{eq:hftraj:srlo}
\Delta N^\hf=-\frac{1}{\Mp^2}\int\left. \frac{V}{V^\prime}\frac{1-\epsilon_1^\hf/3+\epsilon_2^\hf/6}{1-\epsilon_1^\hf/3}\mathrm{d}\phi\right.\, ,
\eeq
where one has used Eqs.~(\ref{eq:hf:eps1}) and (\ref{eq:hf:eps2}) to introduce $\epsilon_1^\hf$ and $\epsilon_2^\hf$. One should notice that the first term of the integrand $V/V^\prime$ actually corresponds to the leading order of the slow-roll trajectory~(\ref{eq:sr:trajlo}). This confirms that when the slow roll is well verified in the horizon-flow parametrization $\epsilon^\hf\ll 1$, the two trajectories are similar.

However, beyond this simple argument, the trajectories' difference is the origin of two subtleties which we now describe, and which biases the horizon-flow analysis.
\subsubsection{$H$-multivaluated trajectories}
Along horizon-flow trajectories, let us first recall that the inflaton field can only vary monotonously. As a consequence, if the complete trajectory is made of several pieces with different signs of $\dot{\phi}$, only one of them can be described by the horizon-flow parametrization, which therefore may be unable to describe the actual outcome of the process. 

Let us illustrate our point on the example of Fig.~\ref{fig:TrajEx}. For the potential displayed in the left panel, the Klein-Gordon equation~(\ref{eq:KG}) is integrated from some initial conditions (specified in the caption), and gives the trajectory displayed in the middle panel (colored lines) in the phase plane $(\phi,\dot{\phi})$. It is made of several pieces : first the inflaton field climbs up the potential (green); then its velocity vanishes and it goes down the potential, it crosses the minimum and it climbs up on the other side (blue); and then its velocity vanishes again and the same thing happens the other way down (red), so on and so forth. In the right panel, the Hubble parameter $H(\phi,\dot{\phi})$ is evaluated along this trajectory with Eq.~(\ref{eq:Friedman}), and  the same color code is adopted. In the middle and right panels, the black dashed line stand for the slow-roll leading order solution, and the arrows indicate in which direction inflation proceeds. One can verify that $H$ always decreases as inflation proceeds.

\begin{figure*}[t]
\begin{center}
\includegraphics[width=\widthsingle]{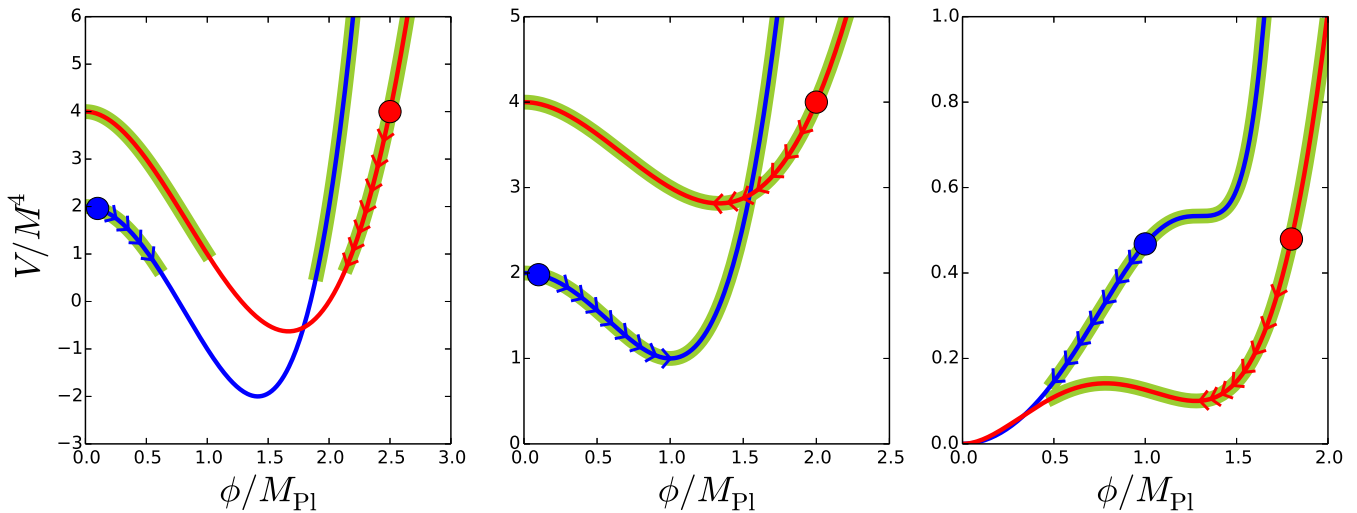}
\caption{Inflationary regimes for a few potential examples (sketch). On each panel, the functional form of $V$ is the same but its coefficients differ (blue and red curves). Regions supporting slow-roll inflation are thickened in green. The horizon-flow trajectory is displayed with the arrows and starts from the location of the disk.}
\label{fig:Traj}
\end{center}
\end{figure*}

In this example it is straightforward to understand why the horizon-flow parametrization cannot describe the entire trajectory: in the right panel, one can check that for a single value of $\phi$, there are several possible values of $H$. The Hubble parameter is therefore multivaluated along the full trajectory, and a single $H(\phi)$ function cannot fully stand for it. 

Since any physical trajectory within the potential eventually approaches the minimum of the potential $\phi=0$, in the spirit of the horizon-flow algorithm [especially case (ii); see section~\ref{sec:hfintro}], this is where the late-time fixed point lies. However, even if the $H$ functions displayed in the right panel of Fig.~\ref{fig:TrajEx} were completed to be defined for all values of the inflaton field, $\phi=0$ would be a late-time fixed point for none of them since it is a minimum of $H$ for none of them.\footnote{Recall that, since $H$ always decreases during inflation, a late-time fixed point is necessarily a minimum of $H(\phi)$.} Therefore, this kind of situation may not be properly described by the horizon-flow parametrization. These aspects are further developed in section~\ref{sec:HFversusSRtraj}
\subsubsection{Inflationary Regime Bias }
In general, a given potential can support inflation in different regimes, but the horizon-flow trajectory is close to the slow-roll one for some of them only. More precisely, letting all the flow parameters vanish beyond some order $M$ sets the functional form of $H$ (hence of $V$), and drawing the initial values $\epsilon_\uin$ of the remaining flow parameters sets the coefficients of these functions as well as the starting point of the numerical integration. Therefore, each time one draws some $\epsilon_\uin$ coefficients, one draws a specific potential $V$ and an initial value $\phi_\uin$ on it. A few examples are displayed in Fig.~\ref{fig:Traj}.

On each panel, the functional form of $H$ from which $V$ is obtained through Eq.~(\ref{eq:VversusH}) is the same but its coefficients are different (blue and red curves). The horizon-flow trajectory is displayed with the arrows and starts from the location $\phi_\uin$ displayed by the disk. On each potential, the green zone denotes where a slow-roll regime of inflation can be supported. One can see that when several slow-roll regimes are possible, $\phi_\uin$ selects out only one of them, such that the other ones are not described. Of course, one can draw other values of $\epsilon_\uin$ so that a similar regime is tracked down (from red to blue case in each panel), but since  $\phi_\uin$ and the coefficients of $V$ are entangled by the choice of $\epsilon_\uin$, this will be done on a different potential. In this sense horizon flow is a biased parametrization of inflation, since once the potential is fixed, only a specific regime out of (possibly) many is worked out. This effect is explicitly exemplified and computed in section~\ref{sec:trajPredictions}.

Finally, it is worth mentioning that when the slow-roll conditions are well verified for $H^\sr$ but not for $H^\hf$, the horizon-flow and slow-roll trajectories are very different; see Eq.~(\ref{eq:hftraj:srlo}). For example, when the second term of Eq.~(\ref{eq:hftraj:srlo}) is negative, the horizon-flow dynamics describes a situation where the inflaton field climbs up its potential. In this case the horizon-flow trajectories have nontrivial instability properties that are further investigated in section~\ref{sec:pathotraj}.

\subsection{A Warm-Up Example: $\boldsymbol{H=1+a\phi}$}
\label{sec:WarmUp}
To briefly summarize the computational program of the following sections, and as an illustrative warm-up, let us consider the simple case where the expansion~(\ref{eq:Hpolynomial}) is truncated at first order; that is, $H/H_0=1+ax$, where $x\equiv \phi/\Mp$. The Hubble parameter is positive provided $x>-1/a$ if $a>0$ and $x<-1/a$ if $a<0$. For simplicity, we only detail the case $a>0$ since $a<0$ is completely symmetrical and can be worked out in exactly the same way. 
In this section we calculate the ratio $r_{16}/(1-\nS)$ predicted by this model, successively making use of the horizon-flow and of the slow-roll setups.

\subsubsection{Horizon-Flow Predictions}
When $a>0$, $H$ increases with $x$; hence from Eq.~(\ref{eq:phidotHprime}) $x$ decreases as inflation proceeds. It stops when $\epsilon_1=2\Mp^2H^{\prime 2}/H^2=2a^2/(1+ax)^2=1$ [see Eq.~(\ref{eq:hf:eps1})], \ie at the location $\xend^\hf$ given by
\beq
\label{eq:hf1i:xend:hf}
\xend^\hf=\sqrt{2}-\frac{1}{a}\, ,
\eeq
where as before the superscript $\mathrm{HF}$ stresses that for now, the calculation is carried out in the horizon-flow framework. The inflationary trajectory is given by Eq.~(\ref{eq:hf:traj}) and the number of $\ee$-folds between the Hubble crossing time of the pivot scale and the end of inflation reads
\beq
\label{eq:hf1i:trajHF}
\Delta N_*^\hf=\frac{x_*^\hf}{2a}+\frac{\left(x_*^\hf\right)^2}{4}-\frac{x_\uend^\hf}{2a}-\frac{{\left(x_\uend^\hf\right)}^2}{4}\, .
\eeq
This trajectory can be inverted, and making use of Eq.~(\ref{eq:hf1i:xend:hf}), one obtains 
\beq
x_*^\hf=-\frac{1}{a}+\sqrt{2+4 \Delta N_*^\hf}\, .
\eeq
Then one needs to plug this expression in the slow-roll parameters $\epsilon_1$ and $\epsilon_2$ expressions, respectively given by Eqs.~(\ref{eq:hf:eps1}) and (\ref{eq:hf:eps2}). Here, since $H^{\prime\prime}=0$, one has $\epsilon_2=2\epsilon_1$. Finally, $r$ and $\nS$ are evaluated. At leading order in slow roll, they are given by Eqs.~(\ref{eq:ns:srlo}) and (\ref{eq:r:srlo}), and one obtains
\beq 
\label{eq:hf1i:hf:lo}
\left. \frac{r_{16}}{1-\nS}\right\vert_\hf^\lo=\frac{1}{4}\, ,
\eeq
which neither depends on $a$ nor on $\Delta N_*$. If one goes up to next-to-leading order in slow roll, $\nS$ and $r$ are respectively given by Eqs.~(\ref{eq:ns:srnlo}) and (\ref{eq:r:srnlo}). Since $H^{\prime\prime}=0$ implies that $\epsilon_3=2\epsilon_1$ in Eq.~(\ref{eq:hf:eps3}), this means that at next-to-leading order in slow roll, one has $r_{16}/(1-\nS)=(1-2\epsilon_{1*})/4$, which yields
\beq
\label{eq:hf1i:hf:nlo}
\left. \frac{r_{16}}{1-\nS}\right\vert_\hf^\nlo=\frac{1}{4} -\frac{1}{2+4 \Delta N_*}\, ,
\eeq
which now mildly depends on $\Delta N_*$ but still not on $a$. For $\Delta N_*\simeq 50$, one obtains $r_{16}/(1-\nS)\simeq 0.245$.

\subsubsection{Slow-Roll Predictions}

Let us now see what the slow-roll setup predicts for this model. Thanks to Eq.~(\ref{eq:VversusH}), the problem consists in studying slow-roll inflation in the potential
\beq
V=3\Mp^2 H_0^2\left(a^2x^2+2ax+1-\frac{2}{3}a^2\right)\, .
\eeq
As before, we restrict ourselves to the case $a>0$. One can see that the potential is definite positive only if $x>-1/a+\sqrt{2/3}$. Therefore there exists a domain, namely $-1/a<x<-1/a+\sqrt{2/3}$, for which $H$ is well defined but not $V$. Fortunately $V(\xend^\hf)>0$ so that this region is never probed, but as explained in sections~\ref{sec:SpecificPot} and \ref{sec:SpecificTraj} this is not the case in general. In the same manner, one can check that $V$ and $H$ both increase with $x$, which means that along the horizon-flow trajectory the inflaton rolls down its potential. However, as mentioned in section~\ref{sec:Traj} and further developed in the following, this is also not always necessarily true.

For what matters now, if one uses Eq.~(\ref{eq:srhierarchySReps1}) to compute $\epsilon_1^\lo$, the end of inflation is determined to happen at 
\beq
\label{eq:hf1i:sr:xend:lo}
\xend^{\sr,\lo}=\sqrt{\frac{7}{6}}+\frac{1}{\sqrt{2}}-\frac{1}{a}\, ,
\eeq
where again the superscript $\sr$ recalls that we are working in the slow-roll framework. The difference with Eq.~(\ref{eq:hf1i:xend:hf}) is not surprising since, here, Eq.~(\ref{eq:srhierarchySReps1}) is used in a regime where, by definition, the slow-roll approximation is not valid anymore.

At leading order in slow roll, the slow-roll trajectory~(\ref{eq:sr:trajlo}) gives rise to
\bea
\label{eq:hf1i:trajSR}
\Delta N_*^{\sr,\lo}&=& \frac{x_*^{\sr,\lo}}{2a}+\frac{\left(x_*^{\sr,\lo}\right)^2}{4}-\frac{x_\uend^{\sr,\lo}}{2a}-\frac{{\left(x_\uend^{\sr,\lo}\right)}^2}{4}
\nonumber\\& &+\frac{1}{3}\ln\left(\frac{1+a\xend^{\sr,\lo}}{1+a x_*^{\sr,\lo}}\right)\, ,
\eea
which resembles the horizon-flow trajectory, but with logarithmic corrections. In practice, Eq.~(\ref{eq:hf1i:trajSR}) needs to be inverted numerically to get $x_*^\sr$. However, to allow comparison with the horizon-flow predictions which do not depend on $a$, let us derive the corresponding slow-roll results in the limit $a\ll 1$. In this case, an approximated formula for the inverted trajectory can be obtained, namely
\bea
x_*^{\sr,\lo}&\underset{a\ll 1}{\simeq}&-\frac{1}{a}+\frac{2\sqrt{2}}{3}\frac{3+\sqrt{21}}{7+\sqrt{21}}+\frac{2}{7+\sqrt{21}}\times
\nonumber\\& &
\sqrt{23+5\sqrt{21}+4\left(14+3\sqrt{21}\right)\Delta N_*}\, .
\eea
The slow-roll parameters at the time of Hubble scale crossing are then obtained by plugging the previous in Eqs.~(\ref{eq:srhierarchySReps1}) and (\ref{eq:srhierarchySReps2}), and $\nS$ and $r$ are computed using Eqs.~(\ref{eq:ns:srlo}) and (\ref{eq:r:srlo}). One obtains a long but explicit expression. For display convenience and again, to allow an easy comparison with the horizon-flow predictions, it can be simplified in the limit $\Delta N_*\gg 1$, and one gets
\beq 
\label{eq:hf1i:sr:lo}
\left. \frac{r_{16}}{1-\nS}\right\vert_\sr^\lo\underset{\underset{\Delta N_*\gg 1}{a\ll 1}}{\simeq}\frac{1}{4}-\frac{7-\sqrt{21}}{96 \Delta N_*}\, .
\eeq
Taking $\Delta N_*\simeq 50$, one obtains $r_{16}/(1-\nS)\simeq 0.2496$. Finally, let us see how these expressions are modified when computing them at next-to-leading order in slow roll. First, if one uses $\epsilon_1^\nlo$ rather than $\epsilon_1^\lo$ to determine $x_\uend$, then one faces a problem since $\epsilon_1^\nlo$ reaches a maximum which is less than $1$ and then becomes negative as inflation approaches its end. This is because the expansion~(\ref{eq:srhierarchySRnloeps1}) does not make sense close to the end of inflation where slow roll is violated. In any case the precise value of $\xend$ does not play a crucial role. Therefore one can safely continue to work with Eq.~(\ref{eq:hf1i:sr:xend:lo}) as far as the location of the end of inflation is concerned. Then, the slow-roll trajectory~(\ref{eq:trajSRnlo}) gives
\bea
\label{eq:hf1i:trajSR:nlo}
\Delta N_*^{\sr,\nlo}&\! \!\!\!\!=\! \!\!\!\! \! & \frac{x_*^{\sr,\nlo}}{2a}+\frac{\left(x_*^{\sr,\nlo}\right)^2}{4}-\frac{x_\uend^{\sr,\nlo}}{2a}-\frac{{\left(x_\uend^{\sr,\nlo}\right)}^2}{4}
\nonumber\\& &\!\!\!\!\!\!\!\!\!\! +\frac{2}{3}\ln\left(\frac{1+a\xend^{\sr,\nlo}}{1+a x_*^{\sr,\nlo}}\right)
\nonumber\\& &\!\!\!\!\!\!\!\! \!\!
+\frac{1}{3}\ln\left[\frac{a^2\left(x_*^{\sr,\nlo}\right)^2+2a x_*^{\sr,\nlo}+1-2a^2/3}{a^2\left(\xend^{\sr,\nlo}\right)^2+2a \xend^{\sr,\nlo}+1-2a^2/3}\right]\, .
\nonumber\\ 
\eea
Again, although this cannot be inverted but numerically, an analytical formula can however be obtained in the limit where $a\rightarrow 0$, which reads
\bea
x_*^{\sr,\nlo}\!\! \!\!&\underset{a\ll 1}{\simeq}&\!\!\! \!-\frac{1}{a}-\frac{4\sqrt{2}}{3}\frac{9+\sqrt{21}}{13+\sqrt{21}}+\frac{2\sqrt{2}}{3\left(13+\sqrt{21}\right)}\times
\nonumber\\& &\!\! \!\!\! \!
\sqrt{205+44\sqrt{21}+3\left(43+9\sqrt{21}\right)\Delta N_*}\, .
\eea
The slow-roll parameters at time of pivot scale crossing are then evaluated at this point but this time using the next-to-leading order expressions~(\ref{eq:srhierarchySRnlostart})--(\ref{eq:srhierarchySRnloend}). Doing so, and again working out the $\Delta N_*\gg 1$ limit for a more convenient comparison of the different results, Eqs.~(\ref{eq:ns:srnlo}) and (\ref{eq:r:srnlo}) give rise to\footnote{Since expressions are consistently worked out at next-to-leading order in slow roll, the $\propto\epsilon^2$ terms in the right-hand side of Eqs.~(\ref{eq:ns:srnlo}) and (\ref{eq:r:srnlo}) are evaluated with the leading order formulas for the slow-roll parameters (\ref{eq:srhierarchySRlostart})--(\ref{eq:srhierarchySRloend}), while the $\propto\epsilon$ terms are evaluated with the next-to-leading order formulas~(\ref{eq:srhierarchySRnlostart})--(\ref{eq:srhierarchySRnloend}).}
\beq
\label{eq:hf1i:sr:nlo} 
\left. \frac{r_{16}}{1-\nS}\right\vert_\sr^\nlo\underset{\underset{\Delta N_*\gg 1}{a\ll 1}}{\simeq}\frac{1}{4}-\frac{7}{888}\frac{13-\sqrt{21}}{\Delta N_*}\, .
\eeq
Taking $\Delta N_*\simeq 50$, one obtains $r_{16}/(1-\nS)\simeq 0.2487$. 

On can see that the difference between both frames predictions at next-to-leading order, Eqs.~(\ref{eq:hf1i:hf:nlo}) and (\ref{eq:hf1i:sr:nlo}) ($\sim 0.004$), is of the same order as the leading order difference Eqs.~(\ref{eq:hf1i:hf:lo}) and (\ref{eq:hf1i:sr:lo}) ($\sim 0.0037$) but does not have the same sign. This confirms that both frames predictions differ by quantities that need to be consistently computed at next-to-leading order in slow roll. 

In this simple toy example, both frames predict similar results and disagree only by subdominant quantities of the order of the percent. However, as will be exemplified in section~\ref{sec:SpecificTraj}, this is not always the case.
\section{Why Horizon Flow Predicts $\boldsymbol{r_{16}/(1-n_\mathrm{\textbf{S}})=1/3}$ or $\boldsymbol{0}$ and Why it Should Not}
\label{sec:SpecificPot}
The computational program of horizon flow sketched in section~\ref{sec:hfintro} oversamples two denser regions in the $(r_{16},\nS)$ plane, namely $r_{16}/(1-\nS)=1/3$ and $r_{16}=0$. As explained before, this can be accounted for by a fixed point analysis, which however singles out slightly different predictions, namely $r_{16}/(1-\nS)=1/2$ and $r_{16}=0$. In this section we puzzle out this mismatch, going beyond a fixed point analysis and explicitly solving the horizon-flow dynamics when the expansion~(\ref{eq:Hpolynomial}) is truncated at second order (which we argue is sufficient). We then exhibit simple horizon-flow models which completely break these relations. This illustrates how the above mentioned denser regions are intimately related to the specific choice of the parametrization~(\ref{eq:Hpolynomial}), and show that they do not hint at intrinsic properties of single-field inflation itself.
\subsection{Elucidating Horizon-Flow ``Predictions''}
\label{sec:hf2i}
The inflationary predictions associated with the models~(\ref{eq:Hpolynomial}) cannot be derived analytically in general for an arbitrarily large value of $M$. However, as we now explain, $M = 1$ already allows us to capture most of the physical effects contained in these models.
Indeed, at leading order in slow roll, one has $r_{16}=\epsilon_1$ and $\nS=1-2\epsilon_1-\epsilon_2$. Looking back at Eqs.~(\ref{eq:hf:eps1})--(\ref{eq:hf:eps2}), these observables involve up to the second derivative in $H(\phi)$ only. Since $H$ is assumed not to vary too much during inflation, it seems reasonable to first neglect higher derivatives, and thus to study the models defined by
\begin{equation}
\label{eq:H:hf2i}
H\left(\phi\right)=H_0\left[1+a\frac{\phi}{\Mp}+b\left(\frac{\phi}{\Mp}\right)^2\right]\, .
\end{equation}
\subsubsection{Inflationary Regimes}
\label{sec:hf2i:regimes:hf}
As before, one denotes $x\equiv\phi/\Mp$.
The $H$ function~(\ref{eq:H:hf2i}) is symmetrical with respect to $x_0=a/\left(2b\right)$, and it is therefore enough to study the inflationary dynamics in the range $x>x_0$ only. If $b>0$, $H$ increases with $x$, and if $0<b<a^2/4$, the Hubble parameter is positive only if $x>x_{H=0}$, where
\begin{equation}
\label{eq:xH0}
x_{H=0}=\frac{\sqrt{a^2-4b}-a}{2b}\, .
\end{equation} 
On the other hand if $b<0$, $H$ decreases with $x$, and is positive only if $x<x_{H=0}$. 

The phase space relation~(\ref{eq:phidotHprime}) implies that $H$ must decrease during inflation. Therefore when $b>0$, the inflaton $\phi$ decreases as inflation proceeds, whereas it increases when $b<0$. Eventually inflation stops when $\epsilon_1=1$. In order to determine when this happens, let us calculate the first slow-roll parameters with Eqs.~(\ref{eq:srhierarchystart})--(\ref{eq:srhierarchyend}). They are given by
\bea
\label{eq:hf2I:srstart}
\label{eq:hf2I:sreps1}
\epsilon_1&\!=&\!2\left(\frac{a+2b x}{1+ax+bx^2}\right)^2,\\
\label{eq:hf2I:sreps2}
\epsilon_2&\!=&\!4\frac{a^2 - 2 b + 2 a b x + 2 b^2 x^2}{\left(1+ax+bx^2\right)^2},\\
\label{eq:hf2I:sreps3}
\epsilon_3&\!=&\!4\left(\frac{a+2bx}{1+ax+bx^2}\right)^2\!\!
\frac{a^2 - 3 b + a b x + b^2 x^2}{a^2 - 2 b + 2 a b x + 2 b^2 x^2}
\, ,\nonumber\\
\label{eq:hf2I:srend}
\eea
and the following slow-roll parameters can be derived in the same manner. The first slow-roll parameter $\epsilon_1$ equals $1$ at
\beq
\label{eq:eps1One}
x_{\epsilon_{1}=1}^\pm=-\frac{a}{2b}+\sqrt{\frac{a^2}{4b^2}+2-\frac{1}{b}}\pm\sqrt{2}\, ,
\eeq
which is defined only when $a^2/4\geq b-2b^2$. This leads to five possible regimes that we now describe one by one.
\begin{samepage}
\begin{itemize}
\item[$(\mathrm{i})$] $b<0$
\end{itemize}
\begin{figure}[H]
\begin{center}
\includegraphics[width=\widthdouble]{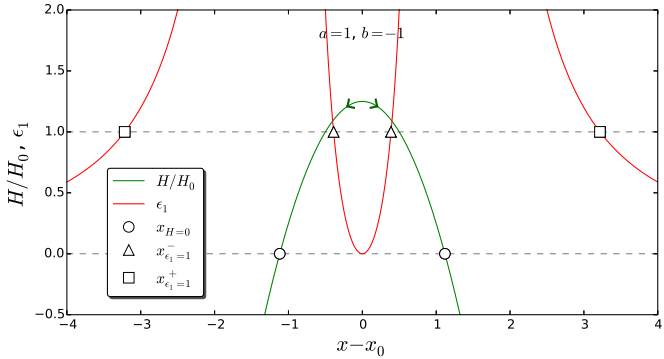}
\caption{Hubble function $H(\phi)$ and first slow-roll parameter $\epsilon_1$ in the case $b<0$ ($a=1$ and $b=-1$).}
\label{fig:Heps1Case1}
\end{center}
\end{figure}
In this case inflation proceeds at $x<x_{H=0}$ for increasing values of $x$ (as denoted by the right arrow in Fig.~\ref{fig:Heps1Case1}), and naturally ends by slow roll
violation since $\epsilon_1$ diverges when $x$ goes to $x_{H=0}$. More precisely, the location at which inflation ends is given by $\xend=x_{\epsilon_1=1}^-$. Note that even if the present analysis is detailed for $x>x_0$ only, symmetrical ranges are displayed in the figures, for illustrative purposes.
\end{samepage}
%\pagebreak
%
\begin{samepage}
\begin{itemize}
\item[$(\mathrm{ii})$] $b\geq 0$ and $a^2>4b$
\end{itemize}
\begin{figure}[H]
\begin{center}
\includegraphics[width=\widthdouble]{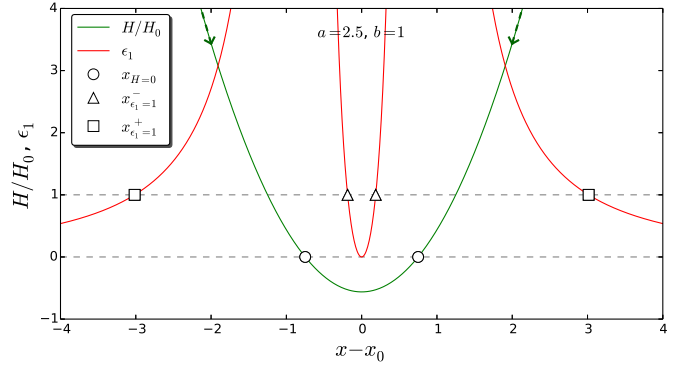}
\caption{Hubble function $H(\phi)$ and first slow-roll parameter $\epsilon_1$ in the case $b\geq 0$ and $a^2>4b$ ($a=2.5$ and $b=1$).}
\label{fig:Heps1Case2}
\end{center}
\end{figure}
Since $b\geq 0$, inflation proceeds for decreasing values of $x$, at $x>x_{H=0}$. Again, $\epsilon_1$ diverges when $x$ goes to $x_{H=0}$ and inflation naturally ends by slow-roll violation, but this time at the location $\xend=x_{\epsilon_1=1}^+$. The dashed lines attached to the arrows mean that inflation actually proceeds at larger values of the field (where $\epsilon_1<1$) than what the position of the arrows indicates.
\end{samepage}
%
%\pagebreak
\begin{samepage}
\begin{itemize}
\item[$(\mathrm{iii})$] $b>0$ and $a^2/4=b$
\end{itemize}
\begin{figure}[H]
\begin{center}
\includegraphics[width=\widthdouble]{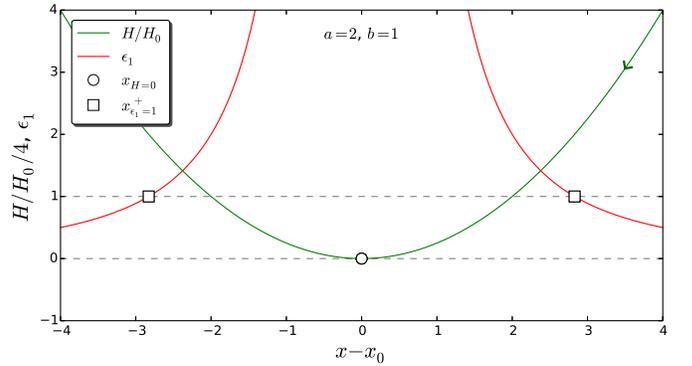}
\caption{Hubble function $H(\phi)$ and first slow-roll parameter $\epsilon_1$ in the case $a^2/4=b$ ($a=2$ and $b=1$).}
\label{fig:Heps1Case3}
\end{center}
\end{figure}
This case is singular since $H$ vanishes only once at $x_0$, where the first slow-roll parameter blows up. Inflation naturally ends at $\xend=x_{\epsilon_1=1}^+$, which simplifies and reads
\beq
\label{eq:xendSing}
\xend=-\frac{2}{a}+2\sqrt{2}\, .
\eeq
\end{samepage}
%\pagebreak
%
\begin{samepage}
\begin{itemize}
\item[$(\mathrm{iv})$] $b>0$ and $b-2b^2\leq a^2/4<b$
\end{itemize}
\begin{figure}[H]
\begin{center}
\includegraphics[width=\widthdouble]{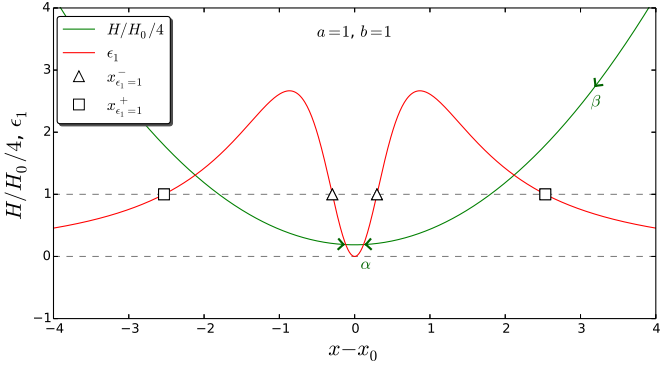}
\caption{Hubble function $H(\phi)$ and first slow-roll parameter $\epsilon_1$ in the case $b>0$ and $b-2b^2<a^2/4<b$ ($a=1$ and $b=1$).}
\label{fig:Heps1Case4}
\end{center}
\end{figure}
In this case $H$ is always positive, and $\epsilon_1$ does not blow up but possesses a maximum that is larger than $1$. This leads to two possible regimes: either inflation starts from $x_\uin<x_{\epsilon_1=1}^-$ and never ends [we call this regime $(\mathrm{iv-}\alpha)$], or it starts from $x_\uin>x_{\epsilon_1=1}^+$ and ends at $\xend=x_{\epsilon_1=1}^+$ [we call this regime $(\mathrm{iv-}\beta)$]. In the $(\mathrm{iv-}\alpha)$ case, $x$ asymptotically approaches the central value $x_0$ and stays there forever.
\end{samepage}
%
%\pagebreak
\begin{samepage}
\begin{itemize}
\item[(v)] $b>0$ and $a^2/4<b-2b^2$
\end{itemize}
\begin{figure}[H]
\begin{center}
\includegraphics[width=\widthdouble]{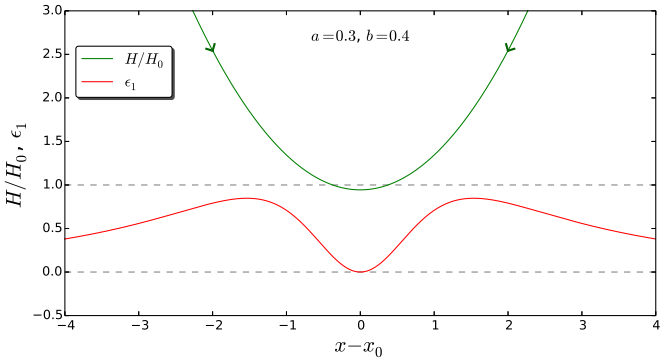}
\caption{Hubble function $H(\phi)$ and first slow-roll parameter $\epsilon_1$ in the case $b>0$ and $a^2/4<b-2b^2$ ($a=0.3$ and $b=0.4$).}
\label{fig:Heps1Case5}
\end{center}
\end{figure}
This last case occurs only if $0<b<1/2$, and is similar to the previous one except that now the maximum value of $\epsilon_1$ is smaller than $1$. Therefore in this situation inflation never ends and $x$ asymptotically approaches the central value $x_0$ where it stays forever.
\end{samepage}
\subsubsection{Inflationary Trajectory}
\label{sec:hf2i:hf:traj}
We can now move on and compute the inflationary trajectory. Integrating the relation~(\ref{eq:dNdphi}) $\dd N=-H/H^\prime \dd\phi/(2\Mp^2)$ between the time $N_*$ when the modes of astrophysical interest today cross out the Hubble radius and the time $N_*+\Delta N_*=N_\uend$ when inflation stops, one obtains
\bea
\Delta N_*^\hf&=&\frac{x_*^2-x_\uend^2}{8}+\frac{a}{8b}\left(x_*-x_\uend\right)
\nonumber\\& &
+\left(1-\frac{a^2}{4b}\right)\frac{1}{4b}\ln\left(\frac{a+2bx_*}{a+2bx_\uend}\right)\, .
\label{eq:traj}
\eea
This trajectory can be inverted to express $x_*$ in terms of $a$, $b$ and $ \Delta N_*$ only. The way it proceeds depends on the case under consideration among the five mentioned above.

When $b<0$, case $(\mathrm{i})$, a first remark is that the number of $\ee$-folds diverges when $x\rightarrow x_0$ and therefore, one is sure to be able to realize a sufficient number of $\ee$-folds. Denoting $X\equiv (a+2 b x)^2/(4b-a^2)$, one obtains $x_*=-a/(2b)-\sqrt{(4b-a^2)X_*}/(2b)$, where
\beq
\label{eq:InvertedTraja}
X_*=W_0\left[X_\uend\exp\left(X_\uend+\frac{32b^2}{4b-a^2}\Delta N_*\right)\right]
\eeq
and where $W_0$ is the $0$ branch of the Lambert function. 

When $b>0$ and $a^2/4>b$, case $(\mathrm{ii})$, one is also sure to be able to realize a sufficient number of $\ee$-folds since $\Delta N$ diverges when $x$ goes to infinity. In this case one obtains $x_*=-a/(2b)+\sqrt{(4b-a^2)X_*}/(2b)$ (notice that the sign of the second term in the right-hand side is different from the case $b<0$), with
\beq
X_*=W_{-1}\left[X_\uend\exp\left(X_\uend+\frac{32b^2}{4b-a^2}\Delta N_*\right)\right]\, ,
\eeq
where $W_{-1}$ is the $-1$ branch of the Lambert function~\cite{Corless:1996zz}. 

When $a^2=4b$, case $(\mathrm{iii})$, the logarithm term in Eq.~(\ref{eq:traj}) vanishes and the trajectory is simply given by
\beq
x_*=-\frac{2}{a}+\sqrt{\frac{4}{a^2}+\xend^2+\frac{4}{a}\xend+8\Delta N_*}.
\eeq
Replacing $\xend$ by Eq.~(\ref{eq:xendSing}), this leads to
\beq
\label{eq:trajSing}
x_*=-\frac{2}{a}+2\sqrt{2}\sqrt{1+\Delta N_*}.
\eeq

Finally when $b>0$ and $a^2/4<b$, cases $(\mathrm{iv})$ and $(\mathrm{v})$, one obtains $x_*=-a/(2b)+\sqrt{(4b-a^2)X_*}/(2b)$ [\ie with the same sign as for case $(\mathrm{ii})$], but with
\beq
\label{eq:InvertedTrajb}
X_*=W_0\left[X_\uend\exp\left(X_\uend+\frac{32b^2}{4b-a^2}\Delta N_*\right)\right]\, .
\eeq
One can check that, as mentioned above, since $\Delta N$ diverges when $x\rightarrow x_0$, an infinite number of $\ee$-folds is realized as $x$ approaches $x_0$ and inflation never ends in the cases $(\mathrm{iii-}\beta)$ and $(\mathrm{iv})$.
\subsubsection{Inflationary Predictions}
\label{sec:hf2i:HF:pred}
\begin{figure*}[t]
\begin{center}
\includegraphics[width=14.cm]{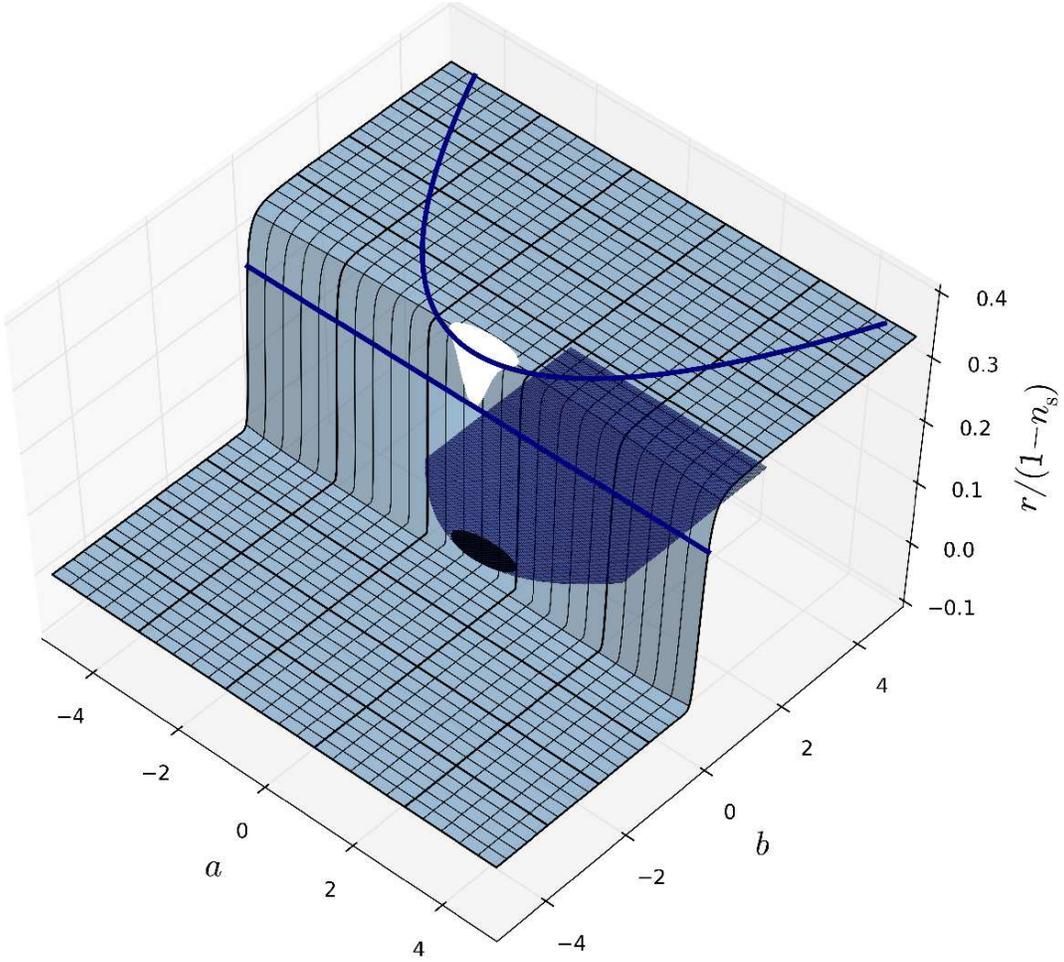}
\end{center}
\caption{Ratio $r_{16}/(1-\nS)$ for the model $H/H_0=1+a x + b x^2$ as a function of $a$ and $b$, with $\Delta N_*=50$. The light blue surface stands for the regimes of inflation where it ends naturally, \ie by slow-roll violation. The ``hole'' in this surface corresponds to case (v) where there is not such a regime. In this case the predictions are calculated at the late-time attractor $x_0$ that yields $r_{16}=0$, which is displayed by the black ellipse behind the light blue surface (and which is the projection of the hole onto the plane $r_{16}=0$). In the same manner, the dark blue surface corresponds to the case $(\mathrm{iv-}\beta)$ where $r_{16}=0$ for the same reason. The blue straight line corresponds to $b=0$ and $r_{16}/(1-\nS)=1/4$, which matches the calculation of section~\ref{sec:WarmUp}, and the blue curved line corresponds to $a^2=4b$, \ie case $(\mathrm{iii})$, for which $r_{16}/(1-\nS)=1/3$ exactly.}
\label{fig:MagicRatio}
\end{figure*}

The physical predictions can now be derived explicitly in terms of $a$, $b$ and $\Delta N_*$, especially the ratio $r_{16}/(1-\nS)$ one is interested in. Making use of  Eqs.~(\ref{eq:ns:srlo}) and (\ref{eq:r:srlo}), at first order in slow roll, it is given by
\beq
\label{eq:ratio:predic}
\frac{r_{16}}{1-\nS}=\frac{\epsilon_1\left(x_*,a,b\right)}{1-2\epsilon_1\left(x_*,a,b\right)-\epsilon_2\left(x_*,a,b\right)}\, ,
\eeq
where the slow-roll parameters $\epsilon(x,a,b)$ are given by Eqs.~(\ref{eq:hf2I:srstart})--(\ref{eq:hf2I:srend}), and the Hubble crossing point $x_*\left(a,b,\Delta N_*\right)$ is given by the formulas detailed above in section~\ref{sec:hf2i:hf:traj}. Obviously it would be straightforward to expand Eq.~(\ref{eq:ratio:predic}) in a (rather long) analytical formula, but one would not learn much doing so. It is instead more instructive to plot the result as a function of $a$ and $b$, which is what is done in Fig.~\ref{fig:MagicRatio}, taking $\Delta N_*=50$ (where we have made sure that different values of $\Delta N_*$ do not modify the result much). 

Let us stress that in the case of never-ending inflation, \ie cases $(\mathrm{iv-}\beta)$ and $(\mathrm{v})$, following the lines of the horizon-flow computational algorithm detailed in section~\ref{sec:hfintro}, the observational predictions are computed at the late-time attractor $x_0$, where $\epsilon_1=0$ and $\epsilon_2\neq 0$; hence $r_{16}/(1-\nS)=0$. This corresponds to the dark blue $(\mathrm{iv-}\beta)$ and black $(\mathrm{v})$ surfaces in Fig.~\ref{fig:MagicRatio}, with the ``hole'' in the blue surface associated with the case $(\mathrm{v})$ where there is no other regime. 

It is also worth mentioning that if $a^2=4 b$, \ie in case $(\mathrm{iii})$, things are particularly simple since Eqs.~(\ref{eq:hf2I:sreps1}) and (\ref{eq:hf2I:sreps2}) combined with Eq.~(\ref{eq:trajSing}) exactly give $\epsilon_{1*}=\epsilon_{2*}=1/(1+\Delta N_*)$; hence $r_{16}/(1-\nS)=1/3$. It is displayed as the blue curved line in Fig.~\ref{fig:MagicRatio}, where one can note the discontinuity in the predictions when $b\rightarrow 0$, this case $(\mathrm{iii})$ being singular. Finally, the blue straight line stands for $b=0$ and $r_{16}/(1-\nS)=1/4$, which corresponds indeed to the calculation of section~\ref{sec:WarmUp}.

One can see that when $b\neq 0$, two asymptotic plateaus are quickly reached. When $b<0$, inflation proceeds close to the maximum of $H$ and in practice, $x_*$ is very close to the local maximum of the potential $x_*\simeq x_0$ where $\epsilon_1=0$ and $\epsilon_2\neq0$. One then typically has $r_{16}=0$, as in the cases $(\mathrm{iv-}\beta)$ and $(\mathrm{v})$. When $b>0$ on the other hand, the asymptotic value of the plateau can be obtained by expanding Eqs.~(\ref{eq:eps1One}), (\ref{eq:InvertedTrajb}), (\ref{eq:hf2I:sreps1}) and (\ref{eq:hf2I:sreps2}) in the limit $b\gg 1$ and $\Delta N_*\gg 1$, and one obtains $r_{16}/(1-\nS)=1/3$, as in the case $(\mathrm{iii})$. These results can be schematically summarized as follows:
\beq
\frac{r_{16}}{1-\nS}\simeq\left\lbrace
\begin{array}{cc}
0&\mathrm{in}\ \mathrm{the}\ \mathrm{cases}\ \mathrm{(i),}\ \mathrm{(iv-}\beta\mathrm{),}\ \mathrm{(v)}\ \\
\frac{1}{3}&\ \mathrm{in}\ \mathrm{the}\ \mathrm{cases}\ \mathrm{(ii),}\ \mathrm{(iii),}\ \mathrm{(iv-}\alpha\mathrm{)}
\end{array}
\right.\, .
\eeq

It is therefore particularly interesting to notice that the typical results found in the literature can simply be interpreted in this framework: $r_{16}=0$ corresponds either to inflation proceeding close to a maximum of $H$ or to a never-ending regime of inflation, while $r_{16}/(1-\nS)=1/3$ is to be associated with a naturally ending inflationary regime where $H$ is not bounded in the far past. This calculation also explains why, as noticed in Refs.~\cite{Kinney:2002qn,Ramirez:2005cy}, the denser regions in the observable plane do not lie exactly at the fixed point $r_{16}/(1-\nS)=1/2$ mentioned in section~\ref{sec:hfintro} but are better described by $r_{16}/(1-\nS)=1/3$.

The mismatch between the numerical results of the horizon-flow computational program and the fixed point analysis is therefore elucidated in this example, and a detailed analysis of the inflationary regimes accounts for the two different typical predictions.
\subsection{Breaking Horizon-Flow ``Predictions''}
\label{sec:simpleextensions}
However, as we shall now see, these ``typical'' predictions are a direct consequence of the parametrization~(\ref{eq:Hpolynomial}), and can easily be broken by other choices of the $H(\phi)$ function.
\subsubsection{$H=1+\alpha \phi^p$}
\begin{figure}[t]
\begin{center}
\includegraphics[width=\widthdouble]{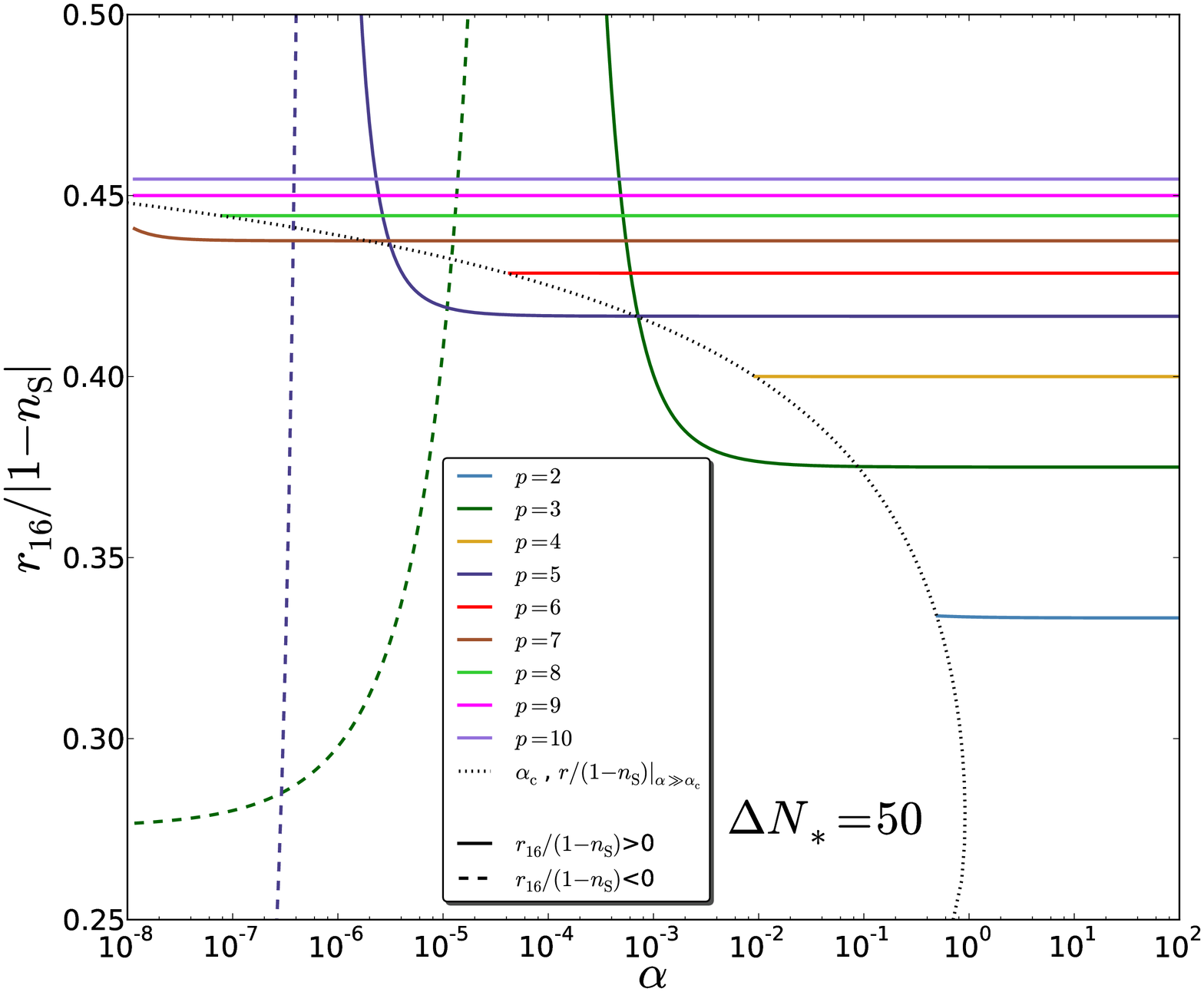}
\end{center}
\caption{Ratio $r_{16}/\vert 1-\nS\vert$ corresponding to the model $H/H_0=1+\alpha x^p$ in the regime of ending inflation, as a function of $\alpha$, for different values of $p$ and with $\Delta N_*=50$. Note that the absolute value of the ratio is displayed (continuous lines when it is positive, and dashed lines when it is negative). When $p$ is even, such a regime exists only when $\alpha$ is larger than $\alpha_\uc$. The black dotted line stands for the $p$-parametrized points $\left(\alpha_\uc,\ r_{16}/(1-\nS)\vert_{\alpha\gg\alpha_\uc}\right)$, where $\alpha_c$ is given by Eq.~(\ref{eq:alphac}) and $ r_{16}/(1-\nS)\vert_{\alpha\gg\alpha_\uc}$ is given by Eq.~(\ref{eq:LimitRatio}).}
\label{fig:MagicRatio2}
\end{figure}
Let us first wonder what would happen if, for some reason, the first terms in the expansion~(\ref{eq:Hpolynomial}) vanish. If $a_1=0$ and $a_2\neq 0$ in Eq.~(\ref{eq:Hpolynomial}), the calculation has already been carried out in section~\ref{sec:hf2i}. Then if $a_1=a_2=0$, the first nonvanishing term of the expansion provides the leading order for $r_{16}$ and $\nS$, for the same reason as that mentioned at the beginning of section~\ref{sec:hf2i}. Let us thus investigate a horizon-flow model of the form $H/H_0=1+\alpha x^p$.

The same detailed analysis as before can be carried out, but here we only summarize the results. Again, when inflation proceeds close to a maximum of $H$, or when inflation never stops and asymptotically reaches a minimum of $H$, one has $r_{16}\simeq 0$. When $\alpha>0$, a regime of naturally ending inflation with unbounded values of $H$ in the far past exists for any $\alpha$ if $p$ is odd, but only if $\alpha$ is larger than some value $\alpha_\uc$ when $p$ is even, given by
\beq
\label{eq:alphac}
\alpha_\uc=2^{-\frac{p}{2}}\left(p-1\right)^{1-p}\, .
\eeq
Indeed, one can check that if $p$ is even and if $\alpha<\alpha_c$, the first slow-roll parameter $\epsilon_1$ is always smaller than $1$. If not, the corresponding location of the end of inflation $\xend$ must be determined numerically in general. One can also check that if $p=2$, the condition $b>1/2$ of case (v) for the model~(\ref{eq:H:hf2i}) matches the value of $\alpha_\uc$ given by Eq.~(\ref{eq:alphac}). The trajectory can be integrated, and one obtains
\beq
\Delta N_*=\frac{1}{2\alpha p}\left(\frac{x_*^{2-p}-\xend^{2-p}}{2-p}+\alpha\frac{x_*^2-x_\uend^2}{2}\right)\, .
\eeq
It is singular when $p=2$, for which the trajectory can directly be read off from Eq.~(\ref{eq:traj}) (taking $a=0$), and it needs to be inverted numerically in general. Doing so, one obtains the value of the field $x_*$ when the modes of astrophysical interest today cross out the Hubble radius. Then, the slow-roll parameters can be evaluated at this point and the corresponding ratio $r_{16}/(1-\nS)$ can be computed. It is displayed in Fig.~\ref{fig:MagicRatio2} as a function of $\alpha$, for integer values of $p$ up to $p=10$, and taking $\Delta N_*=50$ (where we have again made sure that different values of $\Delta N_*$ do not modify the results much).
One can see that when $\alpha$ grows, the ratio $r_{16}/(1-\nS)$ reaches a stationary plateau very quickly, the value of which can be obtained by expanding in $\alpha\gg \alpha_\mathrm{c}$ the previous equations. One then finds $\xend\simeq p\sqrt{2}$ and $\xstar\simeq\sqrt{2p^2+4p\Delta N_*}$. From this it follows that $\epsilon_1^*\simeq p/(2\Delta N_*+p)$ and $\epsilon_2^*\simeq 1/(\Delta N_*+p/2)$, and hence
\beq
\label{eq:LimitRatio}
\left.\frac{r_{16}}{1-\nS}\right\vert_{\alpha\gg\alpha_\uc}\simeq\frac{1}{2\left(1+\frac{1}{p}\right)}\, .
\eeq
The values corresponding to Eqs.~(\ref{eq:alphac}) and (\ref{eq:LimitRatio}) are displayed in Fig.~\ref{fig:MagicRatio2} (black dotted line), where one can check that the matching with the numerical results is very good. On the other hand, if $p$ is odd, the same plateau exists when $\alpha>\alpha_\uc$, where inflation proceeds and stops at $x>0$, but when $\alpha< \alpha_\uc$, inflation can still end naturally when $\epsilon_1=1$ for $x<0$, and a different behavior arises. In this case the ratio $r_{16}/(1-\nS)$ is negative, and it is displayed by the dashed curves in Fig.~\ref{fig:MagicRatio2}.

A continuous set of values for the ratio $r_{16}/(1-\nS)$, including negative ones, is therefore described. One sees that it is enough to allow the cancellation of one or several first terms in the expansion~(\ref{eq:Hpolynomial}) to yield different predictions from $r_{16}=0$ and $r_{16}/(1-\nS)=1/3$. As a consequence, these specific values should not be viewed as generic.
\subsubsection{$r_{16}/(1-\nS)=f$}
One can go even further, adopting a ``reverse engineering'' approach to design a model that gives $r_{16}/(1-\nS)=f$ for any value of $f$. Since $r_{16}/(1-\nS)\simeq 1/(2+\epsilon_2/\epsilon_1)$ at first order in slow roll, any value of $f$ can be reached provided $\epsilon_2=(1/f-2)\epsilon_1$. Plugging Eqs.~(\ref{eq:hf:eps1}) and (\ref{eq:hf:eps2}) in this relation yields a differential equation for $H$, namely $(H^\prime/H)^2=2fH^{\prime\prime}/H$, which can be solved explicitly:
\beq
H\left(x\right)=\left\lbrace
\begin{array}{cc}
H_0\left(1+ax\right)^\frac{2f}{2f-1}&\mathrm{if}\ f\neq\frac{1}{2}\, , \\
 & \\
H_0\,\ee^{ax}&\mathrm{if}\ f=\frac{1}{2}\, , \\
\end{array}
\right.
\eeq
where $a$ is some integration constant. Such $H(\phi)$ functions therefore provide any value $r_{16}/(1-\nS)$ and do not single out any typical inflationary prediction.
\section{Horizon-Flow versus Slow-Roll Trajectories}
\label{sec:SpecificTraj}
As explained in section~\ref{sec:Traj}, the horizon-flow parametrization selects a specific trajectory in phase space among all the solutions of the Klein-Gordon equation~(\ref{eq:KG}) associated with the potential~(\ref{eq:VversusH}) derived from $H$. In this section we first explicitly characterize such trajectories for a few representative examples of inflationary potentials, and compare them with the slow-roll solution which is known as an efficient attractor \cite{Remmen:2013eja} of the inflationary dynamics. This allows us to classify horizon-flow trajectories into three categories, and to highlight that in some cases, horizon flow parametrizes inflation along unstable trajectories where the inflaton climbs up its potential. We then go back to the model~(\ref{eq:H:hf2i}) for which we carry out a complete slow-roll analysis and show that, for a fixed potential, the horizon-flow method is in fact blind to entire inflationary regimes. This introduces a bias in the way it parametrizes inflation. Finally, we numerically investigate, in a reheating consistent manner, the discrepancies in the inflationary predictions obtained from the horizon-flow and the slow-roll trajectories in this model.
\subsection{Horizon-Flow Trajectories in Typical Examples}
\label{sec:HFversusSRtraj}
\begin{figure*}[t]
\begin{center}
\includegraphics[width=\widthsingle]{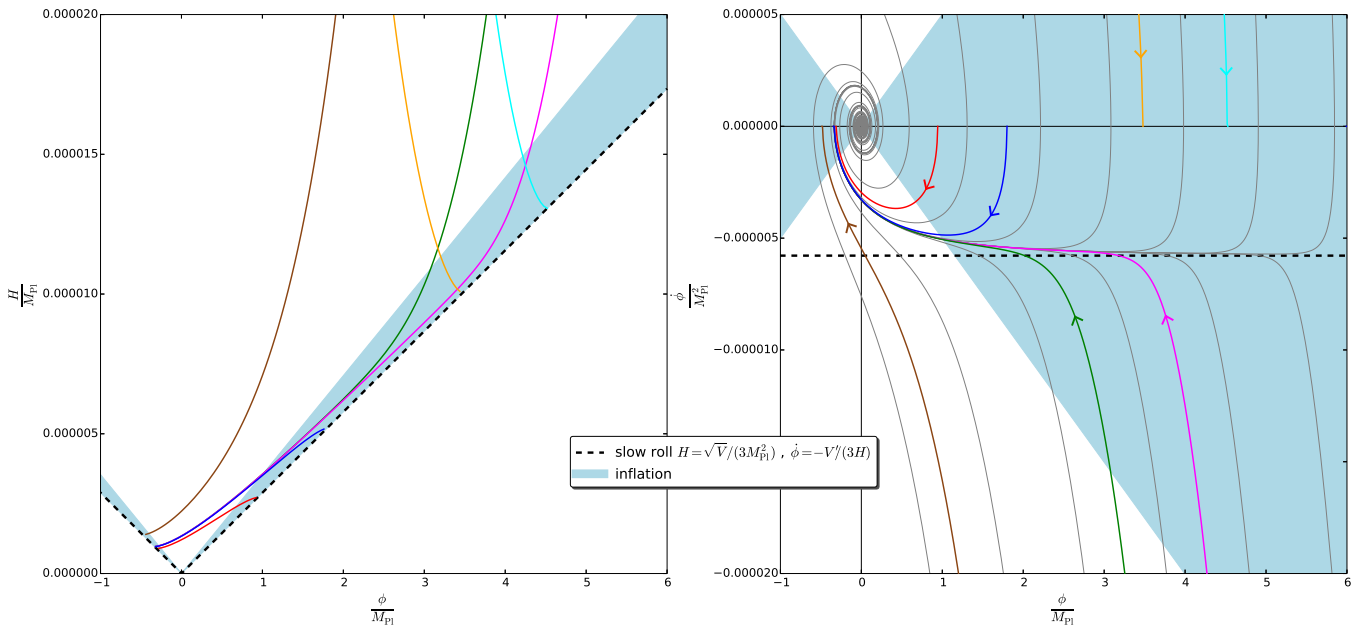}
\end{center}
\caption{Large field inflation, see Eq.~(\ref{eq:lfipot}). The colored lines stand for solutions of Eq.~(\ref{eq:VversusH}) for a few different initial conditions, while the black dashed line represents the slow-roll leading order solution $H^2_{\sr,\lo}\simeq V/3\Mp^2$ (left panel) and $\dot{\phi}_{\sr,\lo}\simeq-V^\prime/(3H_{\sr,\lo})$ (right panel). In the left panel are displayed the $H\left(\phi\right)$ functions, and the corresponding trajectories in phase space $(\phi,\dot{\phi})$ are shown in the right panel. There, grey lines represent numerical solutions of the Klein-Gordon equation~(\ref{eq:KG}) for a few different initial conditions. In both panels, inflation proceeds in the blue surface, defined by $\epsilon_1<1$.}
\label{fig:lfi}
\end{figure*}
\begin{figure*}[t]
\begin{center}
\includegraphics[width=\widthsingle]{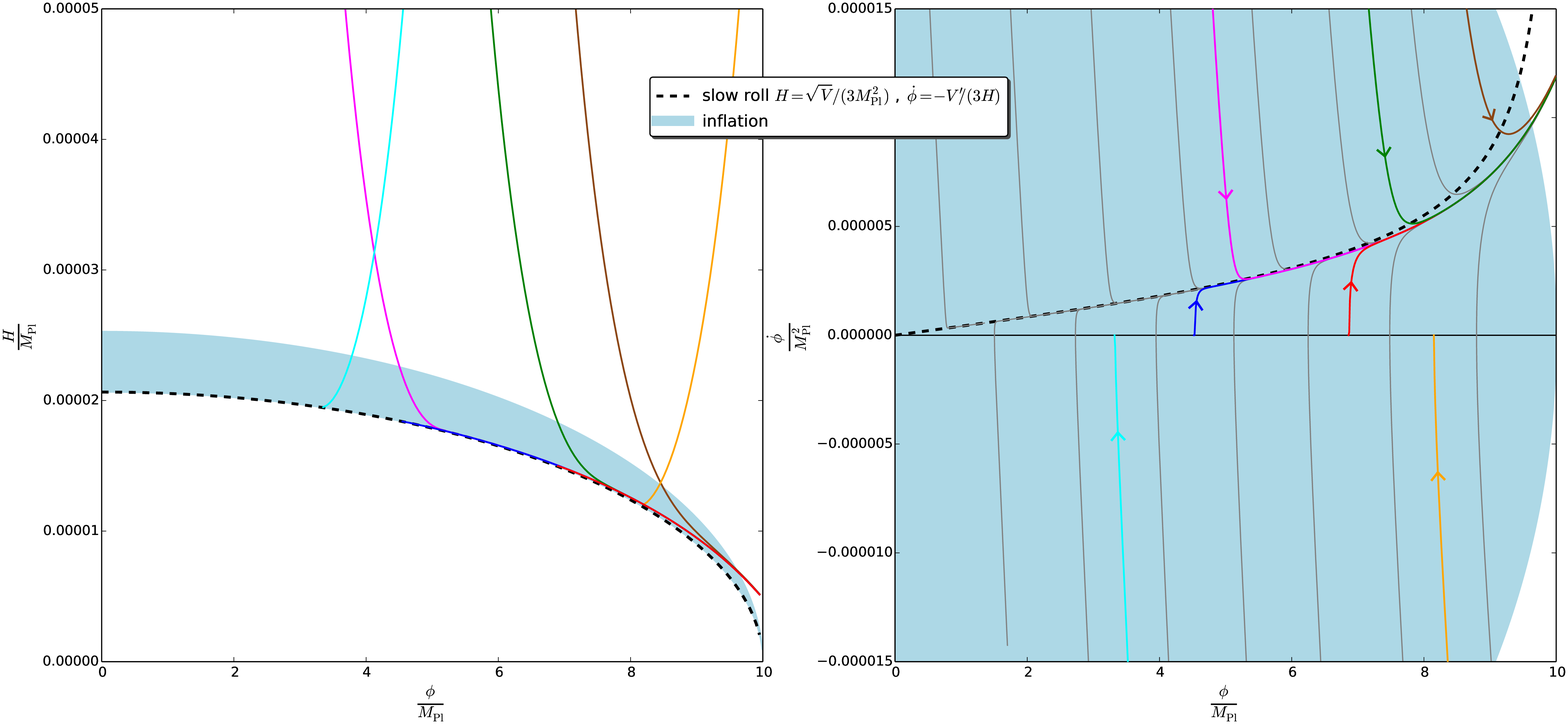}
\end{center}
\caption{Small field inflation, see Eq.~(\ref{eq:sfipot}) for $\mu=\Mp$. The colored lines stand for solutions of Eq.~(\ref{eq:VversusH}) for a few different initial conditions, while the black dashed line represents the slow-roll leading order solution $H^2_{\sr,\lo}\simeq V/3\Mp^2$ (left panel) and $\dot{\phi}_{\sr,\lo}\simeq-V^\prime/(3H_{\sr,\lo})$ (right panel). In the left panel are displayed the $H\left(\phi\right)$ functions, and the corresponding trajectories in phase space $(\phi,\dot{\phi})$ are shown in the right panel. There, grey lines represent numerical solutions of the Klein-Gordon equation~(\ref{eq:KG}) for a few different initial conditions. In both panels, inflation proceeds in the blue surface, defined by $\epsilon_1<1$. }
\label{fig:sfi}
\end{figure*}
\begin{figure*}[t]
\begin{center}
\includegraphics[width=\widthsingle]{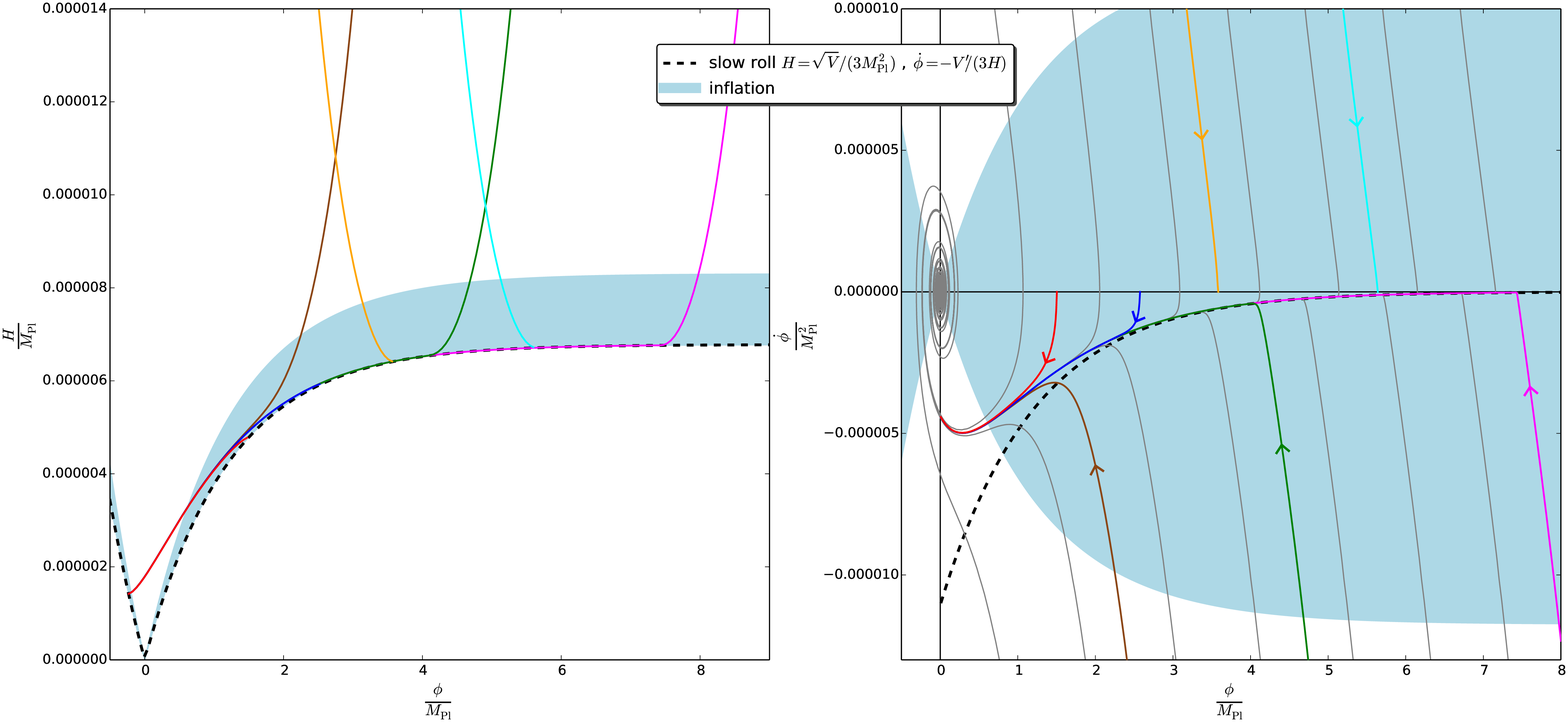}
\end{center}
\caption{Higgs inflation, see Eq.~(\ref{eq:hipot}). The colored lines stand for solutions of Eq.~(\ref{eq:VversusH}) for a few different initial conditions, while the black dashed line represents the slow-roll leading order solution $H^2_{\sr,\lo}\simeq V/3\Mp^2$ (left panel) and $\dot{\phi}_{\sr,\lo}\simeq-V^\prime/(3H_{\sr,\lo})$ (right panel). In the left panel are displayed the $H\left(\phi\right)$ functions, and the corresponding trajectories in phase space $(\phi,\dot{\phi})$ are shown in the right panel. There, grey lines represent numerical solutions of the Klein-Gordon equation~(\ref{eq:KG}) for a few different initial conditions. In both panels, inflation proceeds in the blue surface, defined by $\epsilon_1<1$.}
\label{fig:hi}
\end{figure*}
Let us start again from Eq.~(\ref{eq:VversusH}), $V=3\Mp^2H^2-2\Mp^4H^{\prime 2}$, but let us use it in the opposite way as before: instead of deducing the potential $V(\phi)$ that is associated with a horizon-flow parametrization $H(\phi)$, we now see Eq.~(\ref{eq:VversusH}) as a differential equation giving the $H(\phi)$ functions associated with some potential $V(\phi)$. We shall illustrate our point on several prototypical potential examples. 

In the Schwarz--Terrero-Escalante classification \cite{Schwarz:2004tz} of inflationary potentials, three categories arise depending on the time evolution of the kinetic energy density and its ratio with the total energy density during inflation. The first class is constituted by inflationary models where the kinetic energy and its ratio with the total energy density both increase. It is made of concave potentials. Models belonging to the second class are such that the kinetic energy decreases during inflation, but not its ratio with the total energy density. These models have convex potentials, and a vanishing vacuum constant ($V_\umin=0$). Finally, the third class of inflationary models has a decreasing kinetic energy and a decreasing ratio of the kinetic energy to the total energy density. Its potentials are convex but vacuum dominated ($V_\umin\neq 0$), so that there is no graceful exit to inflation. The third class has been shown~\cite{Martin:2013tda}  to be now ruled out by the most recent observations and we are left with the two first ones. The first class can actually be divided into two subclasses: ``hilltop inflation'' for which inflation proceeds close to a local maximum of its potential, and ``plateau inflation'' where it proceeds along an extended flat portion of it, and where the issue of initial conditions is less acute~\cite{Ijjas:2013vea}.

A prototypical example of hilltop inflation is given by the quadratic ``small field'' potential
\beq
\label{eq:sfipot}
V\left(\phi\right)=M^4\left[1-\left(\frac{\phi}{\mu}\right)^2\right]\, ,
\eeq
where $\mu$ is some mass scale and $M^4$ is an overall energy scale constant, while a common plateau potential is the one of the $f(R)\propto R+R^2$ Starobinsky model \cite{Starobinsky:1980te},
\beq
\label{eq:hipot}
V\left(\phi\right)=M^4\left(1-\ee^{-\sqrt{2/3}\phi/\Mp}\right)^2\, .
\eeq
This potential also appears when inflation is driven by the Higgs field nonminimally coupled to gravity~\cite{Bezrukov:2007ep,GarciaBellido:2011de}, which is naturally the case in curved space-time as produced by quantum fluctuations~\cite{Birrell:1982ix}. This is why in the following we refer to it as the ``Higgs inflation'' model. As a representative of the second class, which is now under observational pressure~\cite{Martin:2013tda}, we take the quadratic potential of ``large field'' inflation
\beq
\label{eq:lfipot}
V\left(\phi\right)=M^4\left(\frac{\phi}{\Mp}\right)^2\, .
\eeq
The shape of these three potentials is displayed below.
\begin{figure}[H]
\begin{center}
\includegraphics[width=\widthdouble]{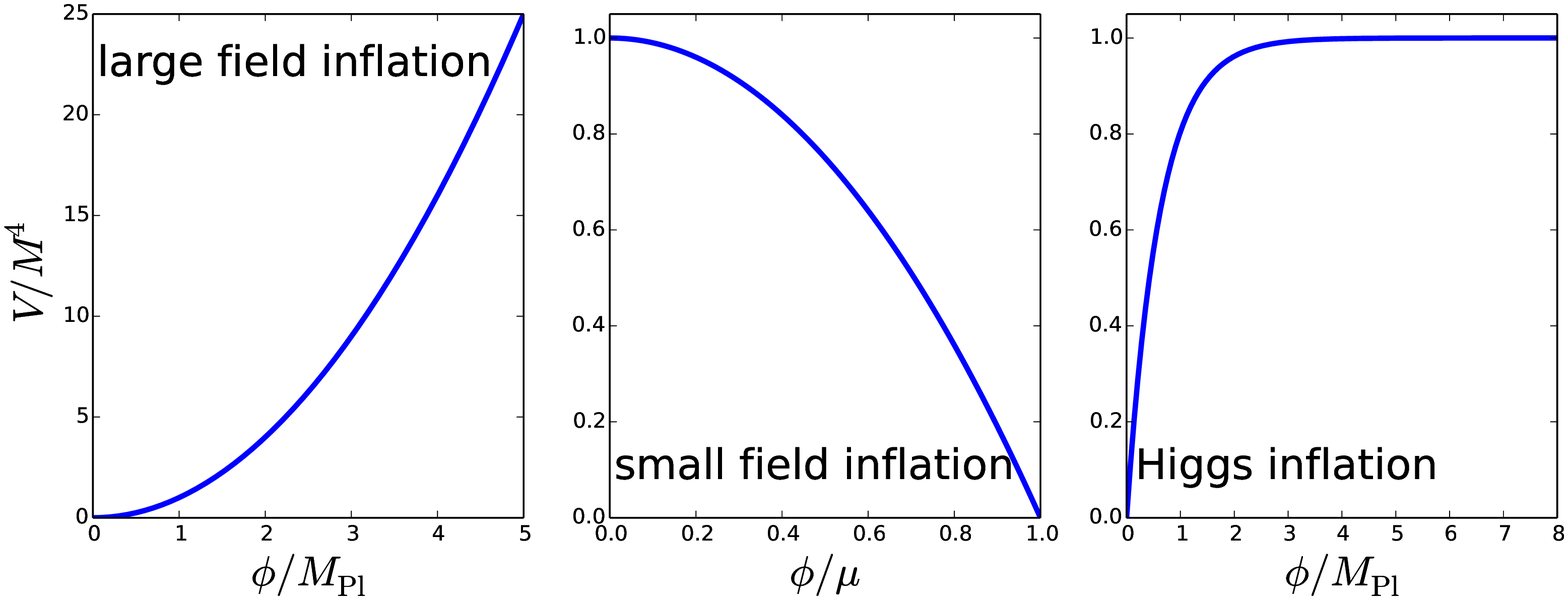}
\label{fig:potExamples}
\end{center}
\end{figure}
For each of these potentials, we solve Eq.~(\ref{eq:VversusH}) numerically and display a few solutions $H\left(\phi\right)$ for different initial conditions (left panels of Figs.~\ref{fig:lfi}, \ref{fig:sfi} and \ref{fig:hi}). The slow-roll leading order solution $H^2_{\sr,\lo}\simeq V/(3\Mp^2)$ is also shown, and the surface where inflation proceeds [defined by $\epsilon_1<1$, or equivalently in the $(\phi,H)$ plane, $H^2<V/(2\Mp^2)$] is also displayed. Let us notice that in general, the solutions $H(\phi)$ can be very different from the slow-roll expected behavior. More precisely, three kinds of solutions $H(\phi)$ are actually obtained. 
\subsubsection{Hubble functions of the first kind}
The first kind is made of Hubble functions which vary with $\phi$ in the same direction as $V$ (\ie $H^\prime V^\prime>0$), and approach the slow-roll solution at late time, \ie when $\phi$ decreases for the large field model, and when $\phi$ increases for the small field and Higgs models. (At very late time, slow roll is violated since inflation ends, and a difference with the slow-roll lowest order solution appears). Such solutions blow up $H^2\gg H^2_{\sr,\lo}$ in the opposite direction, where inflation cannot proceed. Examples of this kind are the magenta, green and brown lines in Figs.~\ref{fig:lfi}, \ref{fig:sfi} and \ref{fig:hi}.  
\subsubsection{Hubble functions of the second kind}
Solutions belonging to the second kind share the same properties, except that at early time they are such that $H^2_{\sr,\lo}<H^2<(V+\dot{\phi}^2_{\sr,\lo}/2)/3\Mp^2$ (where $\dot{\phi}_{\sr,\lo}=-V^\prime/3H_{\sr,\lo}$ is the slow-roll leading order trajectory), and are therefore inflating. However they are not defined in the whole range of possible values for $\phi$ but only in a subinterval. The blue and red lines in Figs.~\ref{fig:lfi}, \ref{fig:sfi} and \ref{fig:hi} are examples of this type.
\subsubsection{Hubble functions of the third kind}
Finally, a third kind is made of solutions which vary with $\phi$ in the opposite direction from $V$ (\ie $H^\prime V^\prime<0$). At late time they approach the lower boundary limit $H^2_{\sr,\lo}$ where they stop being defined, and they blow up at early time where inflation cannot proceed. Examples of this kind are the yellow and cyan lines in Figs.~\ref{fig:lfi}, \ref{fig:sfi} and \ref{fig:hi}. It is important to stress that the condition $H^2>H^2_{\sr,\lo}$ [so that $H^\prime$ is defined in Eq.~(\ref{eq:VversusH})] implies that the functions of the second and third kind are not defined in the full range $\phi>0$. 

Let us see how this translates in terms of phase space trajectories. Once the $H\left(\phi\right)$ function is numerically integrated, the phase space trajectory is directly given by the relation~(\ref{eq:phidotHprime}), $\dot{\phi}=-2\Mp^2H^\prime(\phi)$. The trajectories in phase space $(\phi,\dot\phi)$ corresponding to the $H(\phi)$ functions previously computed are displayed in the right panels of Figs.~\ref{fig:lfi}, \ref{fig:sfi} and \ref{fig:hi} for the three models under study, where the same color code is adopted. The slow-roll leading order trajectory $\dot{\phi}_{\sr,\lo}$ is displayed too, and the inflationary surface defined by $\epsilon_1<1$ [or equivalently $\dot{\phi}^2<V(\phi)$ in the $(\phi,\dot{\phi})$ plane] is also shown. As before, the obtained trajectories can be very different from the slow-roll one. For illustrative purposes, the Klein-Gordon equation~(\ref{eq:KG}) has also been numerically integrated for a few different initial conditions $(\phi_\uin,\dot{\phi}_\uin)$, and the obtained solutions are also displayed. (For the large field model and the Higgs model, the inflaton field oscillates at the bottom of its potential once inflation ends, potentially giving rise to an era of parametric preheating. The oscillations can be noticed in the right panels of Figs.~\ref{fig:lfi} and~\ref{fig:hi}).

The three kinds of Hubble functions listed above can now be easily interpreted. The first kind actually corresponds to an initial overspeed ($\dot{\phi}^2>\dot{\phi}^2_\sr$) that quickly gets damped to the slow-roll attractor solution (brown, green and magenta lines). On the contrary, functions of the second kind correspond to an initial underspeed ($\dot{\phi}^2<\dot{\phi}^2_\sr$) that also quickly reaches the slow-roll attractor solution (blue and red lines). Finally, functions of the third kind are such that the initial velocity of the inflaton is of the opposite sign as the slow-roll one ($\dot{\phi}\dot{\phi}_\sr<0$). In these cases the inflaton initially climbs up its potential (yellow and cyan lines). It is worth stressing that what actually happens in such situations (and as confirmed by the Klein-Gordon numerical solutions) is that the speed of the inflaton decreases and vanishes at some point. Then, the inflaton rolls down its potential and joins the slow-roll attractor. However, the horizon-flow solutions are not able to reproduce these two-step behaviors. Indeed, this means that a single field value $\phi$ is attained several times, with different values of $\dot{\phi}$ (of different signs) each time, and hence different values of $H$. The full inflationary dynamics cannot therefore be described in terms of a single $H(\phi)$ function, and the complete behavior can only be obtained by connecting together a function of the third kind and a function of the second one. This is one of the reasons why some relevant inflationary regimes can be missed by the horizon-flow parametrization.
\subsection{Inflationary Regimes Missed by Horizon Flow}
\label{sec:trajPredictions}
In order to see how these three kinds of Hubble functions are present in the model~(\ref{eq:H:hf2i}) of section~\ref{sec:hf2i}, $H/H_0=1+ax+b x^2$, in this section we turn to the complete slow-roll analysis of this model. In particular, this reveals that, in some cases, the horizon-flow parametrization misses entire inflationary regimes. For an explicit comparison, the predictions provided by both approaches are then also computed in a reheating consistent manner.

The potential associated with the model~(\ref{eq:H:hf2i}) is given by Eq.~(\ref{eq:VversusH}) and reads
\beq
V=3\Mp^2H_0^2\left[\left(1+ax+bx^2\right)^2-\frac{2}{3}\left(a+2bx\right)^2\right]\, ,
\eeq
where we define the overall normalization scale $M^4=3\Mp^2H_0^2$. The slow-roll parameters directly follow from Eqs.~(\ref{eq:srhierarchySRlostart})--(\ref{eq:srhierarchySRloend}) at leading order in slow roll, but here we do not write them down since the expressions are complicated and not very instructive.
\subsubsection{Inflationary Regimes}
\label{sec:hf2iSR:reg}
As for $H$, the potential is symmetrical about $x_0=-a/(2b)$, and therefore it is only necessary to describe it in the $x>x_0$ region.
Following the same logics and notations as in section~\ref{sec:hf2i:regimes:hf}, let us detail the different inflationary regimes that can be supported by the potential.
\pagebreak
%\begin{samepage}
%
\begin{itemize}
\item[$(\mathrm{i})$] $b<0$
\end{itemize}
\begin{figure}[H]
\begin{center}
\includegraphics[width=\widthdouble]{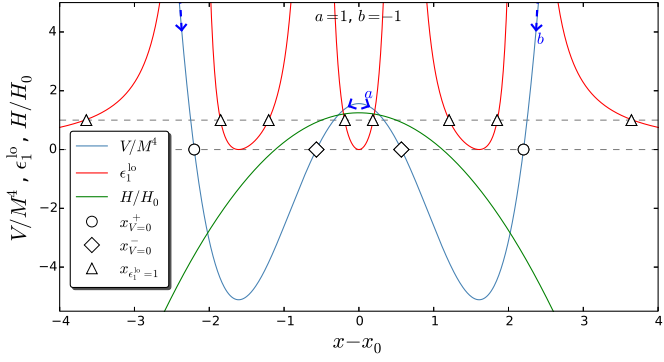}
\caption{Potential $V(\phi)$, first slow-roll parameter $\epsilon_1^\lo$ and $H(\phi)$ function in the case $b< 0$ ($a=1$ and $b=-1$).}
\label{fig:Veps1Case1}
\end{center}
\end{figure}
In this case the potential has a double-well shape. It vanishes at
\beq 
\label{eq:xv0}
x_{V=0}^\pm=-\frac{a}{2b}+\sqrt{\frac{a^2}{4b^2}+\frac{2}{3}-\frac{1}{b}}\pm\sqrt{\frac{2}{3}}\, ,
\eeq
so that inflation takes place either for $x_0<x<x_{V=0}^-$ (where it proceeds from the left to the right) or for $x>x_{V=0}^+$ (where it proceeds from the right to the left). We call these two regimes $(\mathrm{i})\mathrm{-}a$ and $(\mathrm{i})\mathrm{-}b$. The potential is minimal at
\beq
x_{V_\umin}=-\frac{a}{2b}+\sqrt{\frac{a^2}{4b^2}+\frac{4}{3}-\frac{1}{b}}\, ,
\eeq
so that looking back at Eq.~(\ref{eq:xH0}), one has $x_0<x_{V=0}^-<x_{H=0}<x_{V_\umin}<x_{V=0}^+$. Hence there are some values of $x$ (namely $x_{V=0}^-<x<x_{H=0})$ where $H>0$ whereas $V<0$.  However, one can check that $V(x_{\epsilon_1=1}^-)>0$ (where $x_{\epsilon_1=1}^-$ is given by Eq.~(\ref{eq:eps1One}) and is the location where inflation ends in the horizon-flow setup) so that this problematic region is never probed by the horizon-flow trajectory. The equation $\epsilon_1^\lo=1$ can be solved analytically, but the solutions are rather long expressions in $a$ and $b$ that we do not display because they do not add much to this discussion. What is important is that one of these solutions lies in the range $x_0<x<x_{V=0}^-$ and another lies in the range $x>x_{V=0}^+$, so that both regimes $(\mathrm{i})\mathrm{-}a$ and $(\mathrm{i})\mathrm{-}b$ end naturally by slow-roll violation.
 
Finally, it is worth stressing that compared to the horizon-flow $(\mathrm{i})$ case, which only accounts for the ``hilltop'' $(\mathrm{i})\mathrm{-}a$ regime of inflation (\ie inflation occurs close to the maximum of its locally concave potential), the slow-roll analysis reveals the presence of another ``chaotic'' regime (\ie inflation occurs at large field where the potential is convex, and has a graceful exit), case $(\mathrm{i})\mathrm{-}b$, where the $H$ function~(\ref{eq:H:hf2i}) is actually negative and therefore helpless to describe such a regime.
%\end{samepage}
%\pagebreak
%
\begin{samepage}
\begin{itemize}
\item[$(\mathrm{ii})$] $b \geq 0$ and $a^2>4b$
\end{itemize}
\begin{figure}[H]
\begin{center}
\includegraphics[width=\widthdouble]{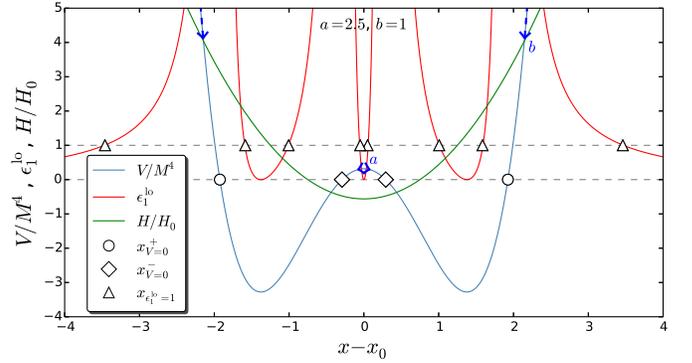}
\caption{Potential $V(\phi)$, first slow-roll parameter $\epsilon_1^\lo$ and $H(\phi)$ function in the case $b\geq 0$ and $a^2>4b$ ($a=2.5$ and $b=1$).}
\label{fig:Veps1Case2}
\end{center}
\end{figure}
As far as the potential is concerned, this case is identical to the previous one $(\mathrm{i})$, and two regimes of inflation are supported, a hilltop one, $(\mathrm{ii})\mathrm{-}a$, and a chaotic one, $(\mathrm{ii})\mathrm{-}b$. However, this time the horizon-flow case $(\mathrm{ii})$ only accounts for $(\mathrm{ii})\mathrm{-}b$, since the $H$ function~(\ref{eq:H:hf2i}) is negative in the $(\mathrm{ii})\mathrm{-}a$ regime.
\begin{itemize}
\item[$(\mathrm{iii})$] $b > 0$ and $a^2=4b$
\end{itemize}
As far as the potential is concerned, this case is not singular and can be described as part of case $(\mathrm{iv})$. However the latter gets divided into several subcases for which the potential behaves differently.
\end{samepage}
\pagebreak
\begin{itemize}
\item[$(\mathrm{iv.1})$] $b > 0$ and $b-2b^2/3\leq a^2/4<b$
\end{itemize}
\begin{figure}[H]
\begin{center}
\includegraphics[width=\widthdouble]{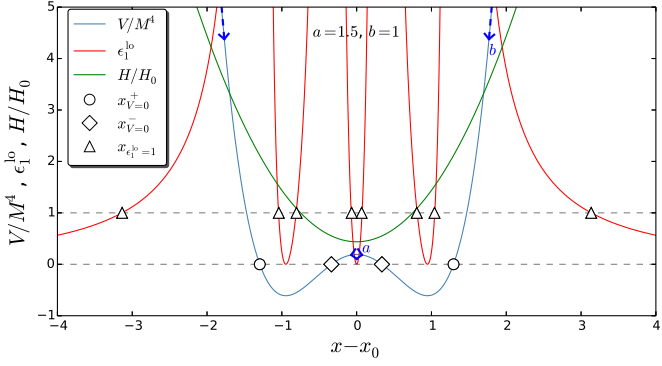}
\caption{Potential $V(\phi)$, first slow-roll parameter $\epsilon_1^\lo$ and $H(\phi)$ function in the case $b > 0$ and $b-2b^2/3\leq a^2/4<b$ ($a=1.5$ and $b=1$).}
\label{fig:Veps1Case41}
\end{center}
\end{figure}
In this case the potential still has the same behavior and supports two regimes of inflation as before, $(\mathrm{iv.1})\mathrm{-}a$ (hilltop) and $(\mathrm{iv.1})\mathrm{-}b$ (chaotic), but now $H(\phi)$ is always positive. This means that contrary to case $(\mathrm{ii})$, the hilltop regime $(\mathrm{iv.1})-a$ can now be described in the horizon-flow setup. However, one should remember that in the corresponding horizon-flow $(\mathrm{iv-}\alpha)$ regime, $x$ decreases as inflation proceeds and approaches $x_0$ where an infinite number of $\ee$-folds is realized. Therefore in this case, the inflaton climbs up its potential and settles over its maximum. In this sense case $(\mathrm{iv-}\alpha)$ is somewhat pathological, and actually corresponds to a Hubble function of the third kind.
\begin{itemize}
\item[$(\mathrm{iv.2})$] $b > 0$ and $b-10b^2/9<a^2/4<b-2b^2/3$
\end{itemize}
\begin{figure}[H]
\begin{center}
\includegraphics[width=\widthdouble]{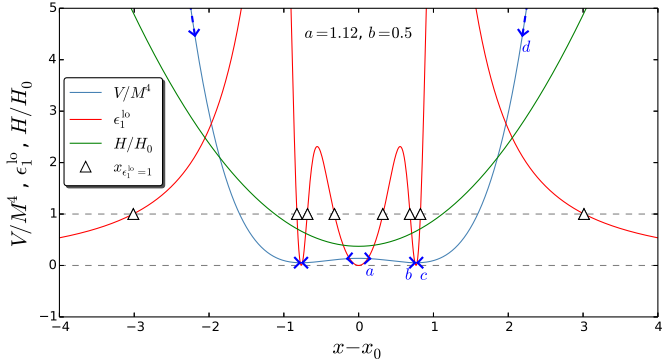}
\caption{Potential $V(\phi)$, first slow-roll parameter $\epsilon_1^\lo$ and $H(\phi)$ function in the case $b > 0$ and $b-10b^2/9<a^2/4<b-2b^2/3$ ($a=1.12$ and $b=0.5$).}
\label{fig:Veps1Case42}
\end{center}
\end{figure}
In this case the potential still has a double well shape but it is positive everywhere. As a consequence there are now four regimes of inflation: a hilltop regime $(\mathrm{iv.2})\mathrm{-}a$ where inflation proceeds from the left to the right close to the maximum of the potential at $x_0$, a vacuum dominated regime $(\mathrm{iv.2})\mathrm{-}b$ where inflation proceeds from the left to the right close to the minimum of the potential at $x<x_{V_\mathrm{min}}$ and does not end by slow-roll violation, another vacuum dominated regime $(\mathrm{iv.2})\mathrm{-}c$ where inflation proceeds from the right to the left close the minimum of the potential at $x>x_{V_\mathrm{min}}$, and a chaotic regime $(\mathrm{iv.2})\mathrm{-}d$ where inflation proceeds from the right to the left at large fields. One should note that depending on the values of $a$ and $b$, it can happen that the local maximum of $\epsilon_1^\lo$ is smaller than $1$, in which case $(\mathrm{iv.2})\mathrm{-}a$ and $(\mathrm{iv.2})\mathrm{-}b$ merge to form a single regime that does not end by slow-roll violation.

Let us compare these regimes to the horizon-flow ones. The regime $(\mathrm{iv.2})\mathrm{-}a$ is described by $(\mathrm{iv-}\alpha)$ during which the inflaton field value decreases, and therefore climbs up its potential and settles over its maximum. The regimes $(\mathrm{iv.2})\mathrm{-}b$ and $(\mathrm{iv.2})\mathrm{-}c$ are not described at all, and $(\mathrm{iv.2})\mathrm{-}d$ corresponds to $(\mathrm{iv-}\beta)$.
%
%\pagebreak
\begin{itemize}
\item[$(\mathrm{iv.3})$] $b > 0$ and $b-4b^2/3<a^2/4<b-10b^2/9$
\end{itemize}
\begin{figure}[H]
\begin{center}
\includegraphics[width=\widthdouble]{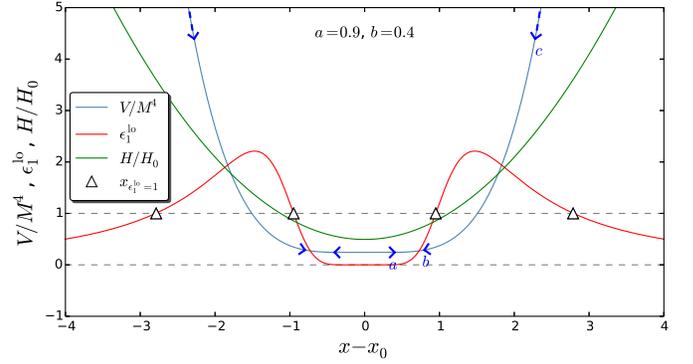}
\caption{Potential $V(\phi)$, first slow-roll parameter $\epsilon_1^\lo$ and $H(\phi)$ function in the case $4(b-4b^2/3)<a^2<4(b-10b^2/9)$ ($a=0.9$ and $b=0.4$).}
\label{fig:Veps1Case43}
\end{center}
\end{figure}
In this case only three regimes exist, namely $(\mathrm{iv.3})\mathrm{-}a$ (hilltop, vacuum dominated), $(\mathrm{iv.3})\mathrm{-}b$ (vacuum dominated) and $(\mathrm{iv.3})\mathrm{-}c$ (chaotic). The first slow-roll parameter $\epsilon_1$ (the one derived from $H$) happens to be greater than $1$ only in the increasing branch of the potential. This means that $(\mathrm{iv.3})\mathrm{-}a$ is described by $(\mathrm{iv-}\alpha)$ during which the inflaton field value decreases (and again climbs up its potential and settles over its maximum), $(\mathrm{iv.3})\mathrm{-}b$ is not described by any horizon-flow regime, and $(\mathrm{iv.3})\mathrm{-}c$ is described by $(\mathrm{iv-}\beta)$.
\pagebreak
\begin{itemize}
\item[$(\mathrm{iv.4})$] $b > 0$ and $b-2b^2<a^2/4<b-4b^2/3$
\end{itemize}
\begin{figure}[H]
\begin{center}
\includegraphics[width=\widthdouble]{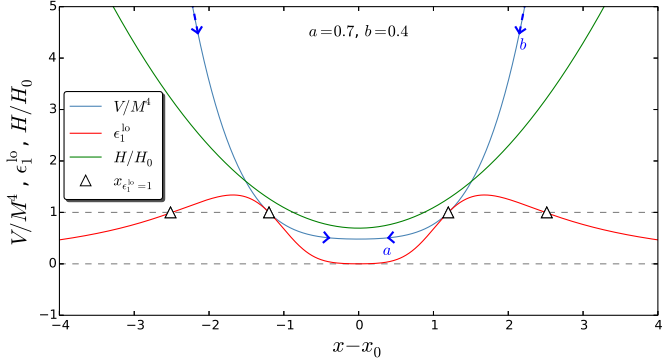}
\caption{Potential $V(\phi)$, first slow-roll parameter $\epsilon_1^\lo$ and $H(\phi)$ function in the case $b > 0$ and $b-2b^2<a^2/4<b-4b^2/3$ ($a=0.7$ and $b=0.4$).}
\label{fig:Veps1Case44}
\end{center}
\end{figure}
In this case the potential is convex everywhere. Two regimes of inflation exist, a vacuum dominated regime $(\mathrm{iv.4})\mathrm{-}a$ where inflation does not end by slow-roll violation and a chaotic regime $(\mathrm{iv.4})\mathrm{-}b$ at large fields. They respectively correspond to $(\mathrm{iv-}\alpha)$ and $(\mathrm{iv-}\beta)$. 
%
%\pagebreak
\begin{itemize}
\item[(v)] $b > 0$ and $a^2/4<b-2b^2$
\end{itemize}
\begin{figure}[H]
\begin{center}
\includegraphics[width=\widthdouble]{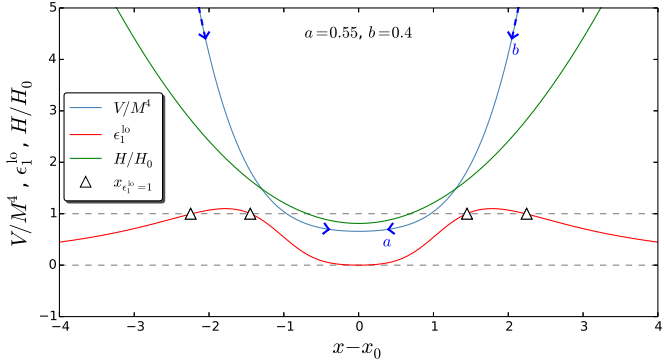}
\caption{Potential $V(\phi)$, first slow-roll parameter $\epsilon_1^\lo$ and $H(\phi)$ function in the case $b > 0$ and $a^2<4(b-2b^2)$ ($a=0.55$ and $b=0.4$).}
\label{fig:Veps1Case5}
\end{center}
\end{figure}
In this case the potential has the same behavior as in the previous case $(\mathrm{iv.4})$. The only difference is that the horizon-flow dynamics only contains a single inflationary regime [remember that in case (v) the first slow-roll parameter $\epsilon_1$, the one derived from $H$, is always smaller than $1$] in which inflation never ends. This corresponds to the regime (v)-$a$, and regime (v)-$b$ is not described in the horizon-flow setup.
However, note that for some values of $a$ and $b$, the local maximum of $\epsilon_1^\lo$ is smaller than $1$ and the two regimes (v)-$a$ and (v)-$b$ merge in a single regime where inflation does not end by slow-roll violation.

To sum up the discussion, we saw that there are some regimes which are supported by the inflationary potential but which are not described by the horizon-flow setup, namely $(\mathrm{i})\mathrm{-}b$, $(\mathrm{ii})\mathrm{-}a$, $(\mathrm{iv.2})\mathrm{-}b$, $(\mathrm{iv.2})\mathrm{-}c$, $(\mathrm{iv.3})\mathrm{-}b$ and $(\mathrm{v})\mathrm{-}a$; and that some hilltop regimes supported by the potential are described by unnatural trajectories in horizon flow where the inflaton climbs up the potential and settles over its maximum, namely $(\mathrm{iv-}\alpha)$ for $(\mathrm{iv.1})\mathrm{-}a$, $(\mathrm{iv-}\alpha)$ for $(\mathrm{iv.2})\mathrm{-}a$, and $(\mathrm{iv-}\alpha)$ for $(\mathrm{iv.3})\mathrm{-}a$. Such trajectories correspond to Hubble functions of the third kind and are somewhat pathological. They are further studied in section~\ref{sec:pathotraj}.

\subsubsection{Inflationary Trajectory}

For now let us move on to the slow-roll trajectory. As explained in section~\ref{sec:Traj}, when slow roll is valid in both the slow-roll and horizon-flow setups, one needs to go up to next-to-leading order in slow roll to consistently compare both frames' predictions. The next-to-leading order slow-roll trajectory~(\ref{eq:trajSRnlo}) can be integrated in our case, and gives rise to
\begin{widetext}
\bea
\Delta N_*^{\sr,\nlo}&=&
\frac{3\left(a^2 - 4 b\right)^2}{16 b^2\left(3 a^2-12b +16 b^2\right)}\ln\left(\frac{a + 2 b x_\uend}{a + 2 b x_*}\right)
%\nonumber\\ & &- 
-\frac{a}{8b}\left(x_\uend-x_*\right)-\frac{1}{8}\left(x_\uend^2-x_*^2\right)
\nonumber\\ & &+
\frac{1}{3}\frac{3a^2-12b+8b^2}{3a^2-12b+16b^2}\ln\left(\frac{3 - 4 b + 3 a x_\uend + 3 b x_\uend^2}{3 - 4 b + 3 a x_* + 3 b x_*^2}\right)
\nonumber\\ & &+
\frac{1}{3}\ln\left[\frac{2a\left(3-4b\right)+2\left(3a^2+6b-8b^2\right)x_\uend+18 a b x_\uend^2+12b^2x_\uend^3}{2a\left(3-4b\right)+2\left(3a^2+6b-8b^2\right)x_*+18 a b x_*^2+12b^2x_*^3}\right]
\nonumber\\ & &- 
\frac{1}{3}\ln\left[\frac{3-2a^2+2a\left(3-4b\right)x_\uend+\left(3 a^2 + 6 b - 8 b^2 \right)x_\uend^2+6 a b x_\uend^3+3b^2x_\uend^4}{3-2a^2+2a\left(3-4b\right)x_*+\left(3 a^2 + 6 b - 8 b^2 \right)x_*^2+6 a b x_*^3+3b^2x_*^4}\right]
\, .
\label{eq:hf2i:trajSR:nlo}
\eea
\end{widetext}
Obviously, this trajectory cannot be inverted analytically and numerical methods must be used in that case.
\subsubsection{Inflationary Predictions}
\begin{figure*}[t]
\begin{center}
\includegraphics[width=\widthdouble]{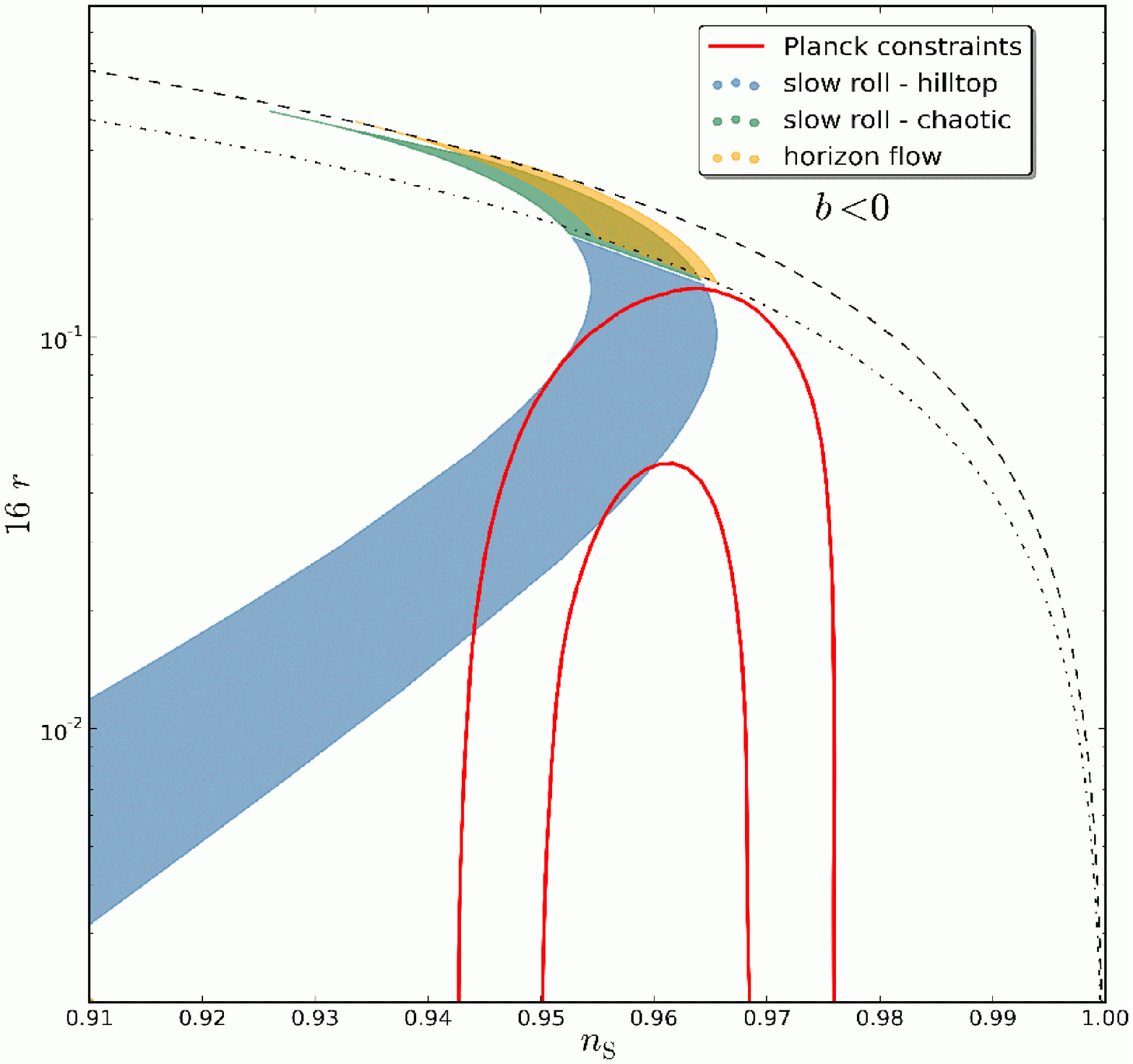}
\includegraphics[width=\widthdouble]{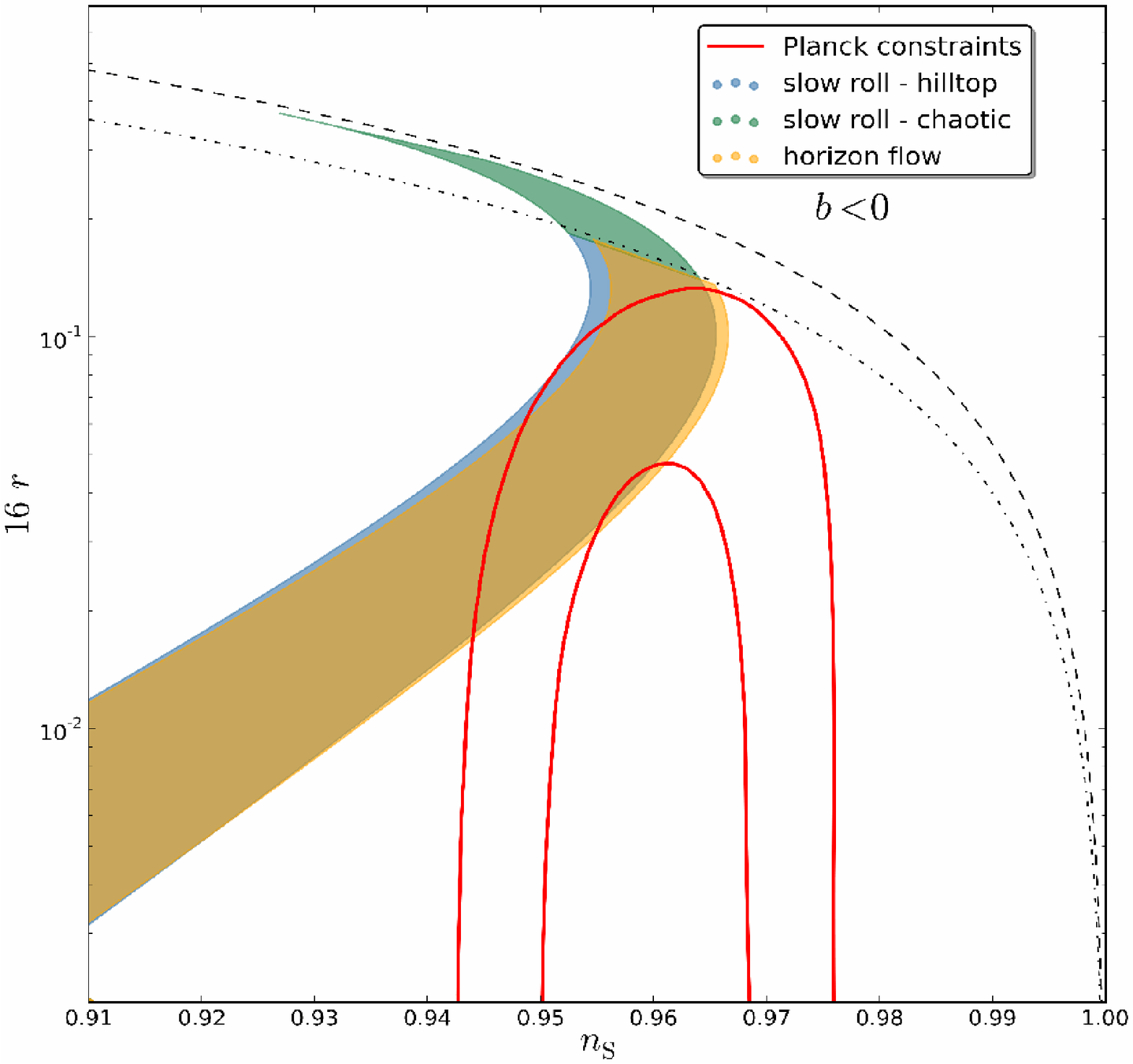}
\end{center}
\caption{Reheating consistent predictions for the model~(\ref{eq:H:hf2i}) in the case $b>0$ (left panel), and $b<0$ (right panel), making use of the horizon-flow setup (yellow area) and the slow-roll one (blue area in the hilltop regime, green area in the chaotic one), and computed at second order in slow roll. The parameters $a$ and $b$ are continuously varied in the ranges $\vert a\vert,\,\vert b\vert\in [10^{-3},10^{3}]$, and $\Delta N_*$ samples values such that $\rho_\mathrm{nuc}<\bar{\rho}_\mathrm{reh}<\rho_\mathrm{end}$ and $\bar{w}_\mathrm{reh}=0$. The two red solid contours are the one- and two-sigma Planck confidence intervals (marginalized over second order in slow roll). The black dashed line stands for $r_{16}/(1-\nS)=1/3$ ($b\gg 1$ limit at first order in slow roll), while the dash-dotted one is for $r_{16}/(1-\nS)=1/4$  ($b\rightarrow 0$ limit at first order in slow roll). One can check that in agreement with Fig.~\ref{fig:MagicRatio}, this ratio $r_{16}/(1-\nS)$ continuously varies in the range $[0,1/3]$.}
\label{fig:HFvsSR}
\end{figure*}
As before, the slow-roll parameters at next-to-leading order, \ie Eqs.~(\ref{eq:srhierarchySRnlostart})--(\ref{eq:srhierarchySRnloend}), are evaluated at the location $x_*$ determined by numerically inverting the trajectory~(\ref{eq:hf2i:trajSR:nlo}). The spectral index and tensor to scalar ratio are then given by Eqs.~(\ref{eq:ns:srnlo}) and (\ref{eq:r:srnlo}). When inflation does not end by slow-roll violation, like in section~\ref{sec:hf2i:HF:pred} we calculate the predictions at the location of the late-time attractor which is the minimum of the potential where $\epsilon_1^\nlo=0$ and $\epsilon_2^\nlo\neq 0$ hence $r_{16}/(1-\nS)=0$.

However, before proceeding with evaluating the slow-roll parameters at the Hubble exit, one must say something about the number of $\ee$-folds $\Delta N_*$. In section~\ref{sec:SpecificPot} the rough estimate $\Delta N_*\simeq 50$ was enough since up to the required precision, the results were not sensitive to $\Delta N_*$. Here however, accurate predictions are needed since we want to compare slow-roll and horizon-flow predictions in the regime where slow roll is valid in both frames. The parameter $\Delta N_*$ must be set consistently with the thermal subsequent history, in particular the reheating stage. Requiring that the mean energy density during reheating $\bar{\rho}_\mathrm{reh}$ be lower than  the energy density at the end of inflation, $\bar{\rho}_\mathrm{reh}<\rho_\mathrm{end}$, and larger than the energy density at, say, the epoch of nucleosynthesis $\bar{\rho}_\mathrm{reh}>\rho_\mathrm{nucl}\,(\simeq$ a few MeV) leads~\cite{Martin:2010kz} to a range of admitted values for $\Delta N_*$. This range depends on the model under consideration and on its parameters,\footnote{The range of admitted values for $\Delta N_*$ also depends on the mean equation of state during reheating $\bar{w}_\mathrm{reh}$, which in Fig.~\ref{fig:HFvsSR} we have set to $\bar{w}_\mathrm{reh}=0$. Changing its value would not modify the discussion.} and needs to be computed numerically. Recently the numerical library \texttt{ASPIC}\footnote{\url{http://cp3.irmp.ucl.ac.be/~ringeval/aspic.html}} has been made public~\cite{Martin:2013tda} and implements such a calculation. Making use of this code, the reheating consistent predictions for the model~(\ref{eq:H:hf2i}) can be worked out, using the horizon-flow formulas of section~\ref{sec:hf2i} on one hand and the slow-roll ones of this section on the other. 

Results are displayed in Fig.~\ref{fig:HFvsSR}, in the $(\nS,r)$ plane, where the Planck mission observational constraints~\cite{Ade:2013rta} are superimposed. The cases $b>0$ and $b<0$ are treated separately since in the horizon-flow parametrization, they correspond respectively to a chaotic regime ($b>0$) where $H$ is not bounded in the far past, and to a hilltop regime where inflation proceeds close to a maximum of $H$ ($b<0$). However, in the slow-roll setup, both regimes can be described in the two cases, and are displayed with different colors to emphasize this crucial point which has important consequences.

For example, one can see that observational constraints favor to a large extent the hilltop regimes with respect to the chaotic ones. Therefore, if, for some reason, one imposes $b>0$, just looking at the horizon-flow parametrization would lead to the biased conclusion that the model is under pressure, whereas a hilltop slow-roll regime actually exists and solves this tension. In this sense, horizon flow may behave as a biased parametrization of inflation.

In passing, one notices that the limit $b\gg 1$ does not give exactly $r_{16}/(1-\nS)=1/3$, mainly because of second order terms that here are taken into account. Finally, as expected, the difference between horizon-flow and slow-roll predictions is rather small inside a given regime.
\subsection{Horizon-Flow Pathological Trajectories}
\label{sec:pathotraj}
In section~\ref{sec:hf2iSR:reg} we made clear that, in the regime $(\mathrm{iv-}\alpha)$, where $4(b-4b^2/3)<a^2<4b$, horizon flow describes inflation along trajectories where the inflaton climbs up its hilltop potential, realizing an infinite number of $\ee$-folds as it approaches the top of the hill.\footnote{In practice, this does not happen since quantum fluctuations start to dominate the inflationary dynamics, which enters a stochastic regime that pushes it away from the potential maximum and connects it with a regular slow-roll phase.}
In this case the Hubble function is of the third kind, according to the typology of section~\ref{sec:HFversusSRtraj}. We end this paper by investigating more these somewhat ``pathological'' trajectories. 

A first interesting remark is that they generalize the ultraslow-roll (USR) scenario~\cite{Kinney:2005vj}. This model is obtained when requiring that the potential is exactly flat $V^\prime=0$ in the Klein-Gordon equation~(\ref{eq:KG}), $\ddot{\phi}/(H\dot{\phi})=-3$, which together with Eq.~(\ref{eq:phidotHprime}) gives a differential equation for $H(\phi)$, namely $\Mp^2 H^{\prime\prime}=3H/2$. One of its solutions is the ultraslow-roll function
\beq
H_\usr=H_0\cosh\left(\sqrt{\frac{3}{2}}\frac{\phi}{\Mp}\right)\, .
\eeq
One can check that Eq.~(\ref{eq:VversusH}) leads to a constant potential $V=3\Mp^2H_0^2$. In this model the inflaton field value approaches $0$ and freezes out there, even if its potential is exactly flat and regardless of its initial value. The slow-roll approximation is never valid since $\epsilon_{2,\usr}=6$, but the model still produces an exactly scale invariant power spectrum, while it produces sizable non-Gaussianities. However, in Ref.~\cite{Martin:2012pe}, it was shown that such a system is unstable and suffers from many physical problems among which is the difficulty to correctly normalize the amplitude of the scalar perturbations to the observed power spectrum. 

The regime $(\mathrm{iv-}\alpha)$ provides a generalized version of ultraslow-roll inflation in the following sense. First, in this case, slow roll is also strongly violated, since $\epsilon_2$ approaches the nonvanishing value
\beq
\epsilon_2\underset{x\rightarrow-a/(2b)}{\longrightarrow} \frac{8b}{1-a^2/(4b)}>\epsilon_{2,\usr}=6
\eeq
as the inflaton reaches the top of the hill, where the last condition comes from the fact that, as recalled above, one is working in the case where $4(b-4b^2/3)<a^2<4b$. Therefore, in some sense, the situation is even worse than in the ultraslow-roll scenario. At the top of the hill $x=-a/(2b)$, inflation still proceeds since $\epsilon_1$ vanishes, but $\epsilon_2$ can be arbitrarily large when $a^2\rightarrow 4b$.

The stability of the inflationary trajectory can also be studied. To do this, let us describe possible deviations from Eq.~(\ref{eq:phidotHprime}) in terms of the modified trajectory
\beq
\label{eq:modifiedHFtraj}
\dot{\phi}=-2\left(1-\delta\right)\Mp^2H^\prime\, .
\eeq
When $\delta=0$, one recovers the horizon-flow trajectory~(\ref{eq:phidotHprime}), but if a small deviation $\delta\neq 0$ is introduced, one is interested in tracking the evolution of its amplitude. First, deriving Eq.~(\ref{eq:modifiedHFtraj}) with respect to time and introducing the Klein-Gordon equation~(\ref{eq:KG}) leads to $\dot{\delta}=[2\Mp^2H^{\prime\prime}(2+\delta)-3H]\delta$. Now, plugging $\dot{\delta}=\delta^\prime\dot{\phi}$, one obtains, at first order in $\delta$, $\delta^\prime=(3/2\,H/H^\prime-2\Mp^2H^{\prime\prime}/H^\prime)\delta$, which has the generic solution
\beq
\delta=\delta_0\frac{H^\prime\left(\phi_0\right)}{H^\prime\left(\phi\right)}\exp\left[\frac{3}{2}\int_{\phi_0}^\phi\frac{H\left(\varphi\right)}{H^\prime\left(\varphi\right)}\dd\varphi\right]\, ,
\eeq
where $\delta_0=\delta(\phi_0)$ is some integration constant. Now, in the model~(\ref{eq:H:hf2i}), as $x$ approaches the top of the potential (and the minimum of $H$) $x_0=-a/(2b)$, this gives rise to
\beq
\label{eq:asymptoticdelta}
\delta\propto\delta_0\left(x+\frac{a}{2b}\right)^{\frac{3}{4}\left(\frac{1}{b}-\frac{a^2}{4b^2}\right)-1}\, .
\eeq
The crucial point is that the exponent appearing in Eq.~(\ref{eq:asymptoticdelta}) is negative as soon as $a^2>4(b-4b^2/3)$, \ie exactly for the case $(\mathrm{iv-}\alpha)$ under study, and $\delta$ blows up at the top of the potential. The horizon-flow trajectory is therefore highly unstable. As a matter of fact, if $\delta_0\neq 0$, either $\phi$ crosses the maximum of its potential and a slow-roll regime possibly occurs on the other half of it ($\delta>0$), or $\dot{\phi}$ vanishes before reaching the top of the potential and a slow-roll regime of inflation then occurs the other way down ($\delta<0$). In any case, the instability of the horizon-flow trajectory~(\ref{eq:phidotHprime}) makes it very unlikely (because the initial condition $\delta_0$ must be fine-tuned to $0$ exactly).

On the contrary, one can check on the right panels of  Figs.~\ref{fig:lfi}, \ref{fig:sfi} and \ref{fig:hi} that the slow-roll solution is a well-behaved attractor, quickly attained from an extended basin of possible initial conditions. More precisely, if initial conditions are such that the kinetic term initially dominates the energy budget of the inflaton field, and that $\dot{\phi}^2\gg\dot{\phi}^2_\sr$, the Klein-Gordon equation~(\ref{eq:KG}) in this ``fast-roll'' limit implies that $\dot{\phi}\propto\ee^{-3N}$. Remembering that inflation starts when $\dot{\phi}^2<V$, this means that the speed of the inflaton is damped to the slow-roll one within a few $\ee$-folds at most, of the order of $\ln\vert\Mp V^\prime/V\vert/3$. This is another reason why the slow-roll setup should be preferred.
\section{Conclusion}
\label{sec:conclusion}
The wide variety of inflationary models makes it tempting to look for model independent approaches for constraining inflationary physics. The horizon-flow strategy has been proposed with exactly this purpose. In this framework generic predictions for the theory of inflation driven by a canonical scalar field have been searched for, and potential reconstruction issues have been investigated. 

The present work showed that the horizon-flow method suffers from a number of flaws, rendering it a somewhat misleading parametrization of inflation.

First, it implicitly relies on phenomenological potentials with no physical justification. Furthermore, instead of choosing priors on potential parameters (which usually stand for physical quantities such as charges, masses, coupling constants, \etc), it samples models from priors defined on unphysical quantities corresponding to initial values of flow parameters.

Second, we have shown that the ``typical'' predictions stemming from this parametrization and that have been noticed in the literature are actually originating from this choice of specific potentials. They can be accounted for analytically going beyond the common fixed point analysis of the problem, and we elucidated the mismatch between the results of this fixed point approach and what was actually numerically obtained by the horizon-flow algorithm. Actually, these predictions turn out to be in direct correspondence with the different regimes of inflation supported by the model. This is why they are not generic features of inflation itself, and explicit examples where they are violated have been provided.

Third, horizon flow implicitly relies on a specific trajectory in phase space among all the solutions of the Klein-Gordon equation associated with the potential it selects. This trajectory is different from the slow-roll one, the latter yet being known as a well-behaved attractor of inflationary dynamics. At first sight, this only leads to slow-roll suppressed discrepancies in the predictions between both frames, which we have computed. More importantly however, we have found that for a given potential, entire regimes of inflation are missed by the horizon-flow approach, which therefore introduces a bias in the analysis. Interestingly enough, in these missed regimes, the horizon-flow trajectory can be highly unstable (corresponding to an inflaton climbing up its potential and asymptotically approaching its local maximum), and provides a generalized and worsened version of ultraslow-roll inflation.

For these reasons, we conclude that even if convenient, studying inflation along the lines of the horizon-flow program can lead to biased or even inexact results.
\acknowledgments
I would like to warmly thank J.~Martin, P.~Peter and C.~Ringeval for careful reading of the manuscript and very useful comments.
\bibliography{biblio}

\thispagestyle{plain}
\begingroup
\parindent 1pt
\parskip 2ex
%\def\enotesize{\normalize}

%\theendnotes

\endgroup

\end{document}